\documentclass[10pt,amsmath,amssymb,nofootinbib,twoside,twocolumn,superscriptaddress,floats,floatfix,aps,prd,preprintnumbers]{revtex4-2}
\usepackage[british]{babel}
\usepackage{xcolor}
\usepackage{booktabs}
\usepackage{epsfig}
\usepackage{latexsym}
\usepackage{amssymb}
\usepackage{amsmath}
\usepackage{graphics}
\usepackage{tabularx}
\usepackage{cancel}
\usepackage{hyperref}
\hypersetup{
    colorlinks=true,
    linkcolor=red,
    citecolor=blue,
    filecolor=black,
    urlcolor=black,
    pdfauthor={D. Brundu, M. Cadoni, P. Olla, M. Oi, A. P. Sanna},
    pdftitle={Atmospheric Newtonian noise modeling for third-generation gravitational wave detectors}
}
\usepackage[capitalise]{cleveref}
\usepackage[utf8]{inputenc}
\usepackage[normalem]{ulem}
\def\beq#1\eeq{\begin{align}#1\end{align}}
\def\d{{\rm d}}
\def\im{{\rm i}}

\def\ex{{\rm e}}
\def\x{{\bf x}}

\def\k{{\bf k}}
\def\U{{\bf U}}

\def\flux{\mathcal{E}}

\def\smalU{{\scriptscriptstyle U}}
\def\PBL{{\scriptscriptstyle PBL}}
\def\CHI{C^{\text{HI}}}
\def\CHIT{C^{\text{HI,T}}}
\def\CT{C^{\text{T}}}
\def\LO{L_{\text{O}}}

\begin{document}
\newcommand{\UniCa}{\affiliation{Dipartimento di Fisica, Universit\`a di Cagliari, Cittadella Universitaria, 09042 Monserrato, Italy}}
\newcommand{\INFN}{\affiliation{I.N.F.N, Sezione di Cagliari, Cittadella Universitaria, 09042 Monserrato, Italy}}
\newcommand{\ISACCNR}{\affiliation{ISAC-CNR, Sezione di Cagliari, Cittadella Universitaria, 09042 Monserrato, Italy}}
\title{Atmospheric Newtonian noise modeling for third-generation gravitational wave detectors}

\author{D.~Brundu}
\email{davide.brundu@ca.infn.it}
\INFN

\author{M.~Cadoni}
\email{mariano.cadoni@ca.infn.it}
\INFN\UniCa

\author{M.~Oi}
\email{mauro.oi@ca.infn.it}
\INFN\UniCa

\author{P.~Olla}
\email{olla@dsf.unica.it}
\INFN\ISACCNR

\author{A.~P.~Sanna}
\email{asanna@dsf.unica.it}
\INFN\UniCa

\begin{abstract}
    The sensitivity and the frequency bandwidth of third-generation gravitational-wave (GW) detectors are such that the Newtonian noise (NN) signals produced by atmospheric turbulence could become relevant. We build models for atmospheric NN that take into account finite correlation times and inhomogeneity along the vertical direction, and are therefore accurate enough to represent a reliable reference tool for evaluating this kind of noise. We compute the NN spectral density from our models and compare it with the expected sensitivity curve of the Einstein Telescope (ET) with the xylophone design. The noise signal decays exponentially for small values of the frequency and the detector's depth, followed by a power-law for large values of the parameters. We find that, when the detector is built at the earth's surface, the NN contribution in the low-frequency band is above the ET sensitivity curve for strong wind. Building the detector underground is sufficient to push the noise signal under the ET sensitivity curve, but the decrement is close to marginal for strong wind. In light of the slow decay  with depth of the NN, building the detector underground could be only partially effective as passive noise mitigation.
\end{abstract}

\preprint{ET-0126A-22}

\maketitle
\section{Introduction}
The first direct observation of gravitational waves (GW) performed in the last years by the LIGO--Virgo collaboration \cite{LIGOScientific:2016aoc,VIRGO:2014yos,LIGOScientific:2016sjg,LIGOScientific:2017vwq,LIGOScientific:2017ync,LIGOScientific:2017zic,LIGOScientific:2016lio,LIGOScientific:2019fpa,LIGOScientific:2020ibl,LIGOScientific:2021qlt} represents  a milestone for fundamental physics and astrophysics. The direct detection of the GW signals generated by coalescing objects like neutron-stars or black-hole binaries has not only provided a striking confirmation of Einstein's General Relativity in the strong-field regime, but has also started the new era of multi-messenger astrophysics. The international network of
second-generation GW detectors has been further enhanced with the joining of the KAGRA detector in 2020 \cite{Somiya:2011np,KAGRA:2013rdx,KAGRA:2020tym}. The currently operating GW detectors use extremely sensitive Michelson interferometers and have a sensitivity band ranging from $10$ Hz to $10$ kHz.  

Third-generation GW detectors like the Einstein Telescope (ET) \cite{Punturo:2010zz} and Cosmic Explorer (CE) \cite{LIGOScientific:2016wof} have been proposed to fully open the emerging field of GW astrophysics and cosmology \cite{Sathyaprakash:2012jk,Maggiore:2019uih}. Their goal is to improve the sensitivity by a factor of $10$ and push the observation band down to $1$ Hz. These improvements are motivated by numerous scientific reasons \cite{Maggiore:2019uih}. Black-hole mergers could be observed at higher redshift and mass; the inspiral phase could be detected earlier allowing for a better multi-messenger investigation of the source; possible quantum gravity effects, e.g. quantum hair(s) for black holes, could be detected in the ringdown phase. These are just a few examples of the scientific relevance of third-generation GW detectors.

Improvements in sensitivity and frequency bandwidth pose formidable challenges due to the impact of various noise sources. This is not only because the enhancement of the sensitivity of a factor of $10$ may push the latter down close to the noise floor, but also because we need better modeling of noise in the lowest part of the frequency band. In fact, at few Hz, the major limitations are expected to come from the gravitational fluctuations, also called gravitational gradient noise or Newtonian Noise (NN) \cite{Har2019,Harms:2022jth}.
NN has two main contributions coming from seismic fields and atmospheric perturbations. While detailed estimations of seismic NN have been performed--also by modeling seismic sources (for a comprehensive review, see, e.g. Ref. \cite{Har2019} and references therein)--atmospheric NN is instead poorly understood. Our understanding of the atmospheric contribution to the NN in GW detectors has, until now, been essentially based on the works of Saulson \cite{Saulson:1984yg} and Creighton  \cite{Creighton:2000gu} (see also Ref.~\cite{Cafaro:2009mu} for an analysis of the contributions coming from pressure fluctuations produced in turbulent flow and Ref.~\cite{Cafaro:2009mu} for an analysis of acoustic NN). Saulson considered the effect of acoustic pressure waves. Conversely, Creighton investigated the contributions to NN of temperature perturbations, transient atmospheric shocks  and sound waves generated by colliding objects.
This lack of interest in the atmospheric contributions to the NN is motivated by the fact that estimations predicting NN in the $10$ Hz region are several orders of magnitudes below the sensitivity curve of second-generation GW interferometers \cite{Har2019}.  

The situation changes drastically when one considers third-generation GW detectors. In this case, the NN prediction derived from modeling wind-advected temperature fluctuations, which are the dominant sources of atmospheric density fluctuations, cannot be extended below $10 \ \rm Hz$ without modifying the models \cite{Creighton:2000gu}. 
This is because the two basic assumptions adopted in \cite{Saulson:1984yg,Creighton:2000gu,Har2019} for the estimation of NN generated by temperature fluctuations,
namely, the quasi-static approximation and the homogeneity and isotropy hypothesis, are expected to fail in that frequency band.

A second crucial issue is that some third-generation GW detectors are planned to be built underground. Although it is generally qualitatively true that underground construction of the interferometers will represent passive mitigation of both seismic and atmospheric NN, quantitative results for the dependence of the noise level on the detector depth  are not presently available. Again, this is because atmospheric NN modeling in the past was oriented towards second-generation GW-detectors, i.e. detectors built on the earth's surface.

The main purpose of this paper is to improve the modeling of NN generated by wind-advected temperature fluctuations, to make it a reliable reference tool for the evaluation of NN for third-generation GW detectors.
This is going to be a necessary ingredient not only for a general preliminary estimation of the noise, but also for the successive, detailed evaluation of NN by numerical simulation of atmospheric flows in realistic conditions. This is mainly because current atmospheric codes have a grid scale that is typically well above that of the fluctuations expected to contribute to NN.

We will improve the quasi-static and homogeneous models of \cite{Creighton:2000gu} by working in two different directions. Firstly, we will go beyond the former approximation by building models in which the NN generated by the decay of the vortices is taken into account. Secondly, we will fully take into account the fact that atmospheric turbulence is strongly inhomogeneous along the vertical direction, and therefore it cannot be modeled \emph{a priori} within a homogeneous-isotropic (HI) turbulence framework. We will find that, whenever vortex time-decay dominates over the wind-advection component, the NN power spectra are characterized by a power-law behavior. This result is only weakly dependent on the specific form of the time correlations of the turbulence, except for its scaling properties. This is fully expected given the multi-scale behavior of turbulent phenomena. On the other hand, the power-law regime of NN spectra has not been previously found in HI models of frozen turbulence \cite{Creighton:2000gu}, and its possible impact on GW detectors was therefore completely overlooked.

We will then apply the results of our models to assess the impact of temperature-fluctuations-induced NN on the planned ET detector. For this purpose, we will compare the numerical NN power spectrum obtained from our models with the expected sensitivity curve of the ET with the xylophone design (ET-D configuration). We will also discuss the dependence of the power spectrum on the physical parameters of our model, putting a particular emphasis on the dependence on the detector depth $r_0$. In the frequency range of interest, we will find that the noise, as a function of $r_0$, decays exponentially for small values of this parameter. A $1/r_0^2$  scaling sets in at relatively large values of $r_0$. This scaling behavior turns out to be generally true  whenever the effect of wind advection is negligible and $r_0$ is large. This implies that building the detector underground could not provide sufficient passive mitigation of atmospheric NN.

The structure of the paper is as follows. In \cref{sectatmnn} we discuss in general terms the modeling of NN from atmospheric temperature fluctuations beyond the quasi-static approximation.
In \cref{sectturbboundlay} we briefly discuss the main features of turbulence in the planetary boundary layer (PBL). We model the NN generated by wind-advected HI turbulence in \cref{sectNNhom}.
In \cref{sec:WallTurbulence} we build a realistic model for NN generated by turbulence in the PBL.
In \cref{sectresults} we present the power spectra derived from our models, for selected values of the parameters, compare them with the ET-D sensitivity curve and discuss our results.
In \cref{sectconclusions} we draw our conclusions. We leave technical details to the appendices.

\section{Atmospheric Noise from Temperature Fluctuation}
\label{sectatmnn}

Density perturbations $\delta \rho(\mathbf{r},t)$ in the air caused by temperature fluctuations $\delta T(\mathbf{r},t)\equiv\tilde T(\mathbf{r},t)$ are a major source of atmospheric NN for GW detectors. The heat in the atmosphere generates convective turbulence, mixing pockets of cold and warm air at all length-scales down to the millimeter. Indicating with $\bar T$ and $\bar \rho$ the mean temperature and density of air, respectively, from the ideal gas law at constant temperature we get $\delta \rho(\mathbf{r},t)=-(\bar\rho/\bar T)\tilde T(\mathbf{r},t)$. Density perturbations generated by temperature  fluctuations are typically several orders of magnitude larger than those generated by pressure perturbations, which disperse in the atmosphere in the form of infrasound waves. The gravitational acceleration perturbation $\delta \mathbf a(\mathbf{r}_0,t)$ produced on the test-mass of the detector located at $\mathbf{r}_0$ at time $t$ is given by
\begin{equation}
\delta\mathbf a(\mathbf{r}_0,t)= - \alpha\int dV \frac{\tilde T(\mathbf{r},t)}{|\mathbf{r}-\mathbf{r}_0|^3}(\mathbf{r}-\mathbf{r}_0),
\label{strain}
\end{equation}
where $\alpha= G \bar\rho/\bar T$ is the conversion factor from temperature to acceleration fluctuations in the detector. The acceleration fluctuation given by \cref{strain} must then be projected onto the detector-arm direction in order to obtain the strain. 

Physically, the time variation of the acceleration $\delta \mathbf a(\mathbf{r}_0,t)$ is the result of two different effects: the decay of vortices and their transport by the wind with average velocity $\mathbf U$ past the detector.
The approach adopted in \cite{Creighton:2000gu,Har2019}, which approximates turbulence as a frozen field, is based on the hypothesis that  the vortex decay time-scale (identified with the eddy turnover time) is much larger than the typical time spent by the vortex in the vicinity of the detector \cite{Creighton:2000gu}. In this approximation, the main contribution to \cref{strain} is due to the effect of frozen temperature fluctuations transported by the wind near the detector. 
As explained in the introduction, it is commonly believed that this approximation breaks down at time-scales larger than $10$ seconds \cite{Creighton:2000gu,Har2019}. However, this is a rather intricate point, which deserves to be carefully analyzed. Indeed, the frozen turbulence limit has to be defined in terms of both the decay time of temperature correlations in the reference frame of the wind and the effective time the turbulent structure is perturbing the detector. We will discuss this issue in detail in \cref{ftl}.

\begin{figure}
\begin{center}
\includegraphics[width=\columnwidth]{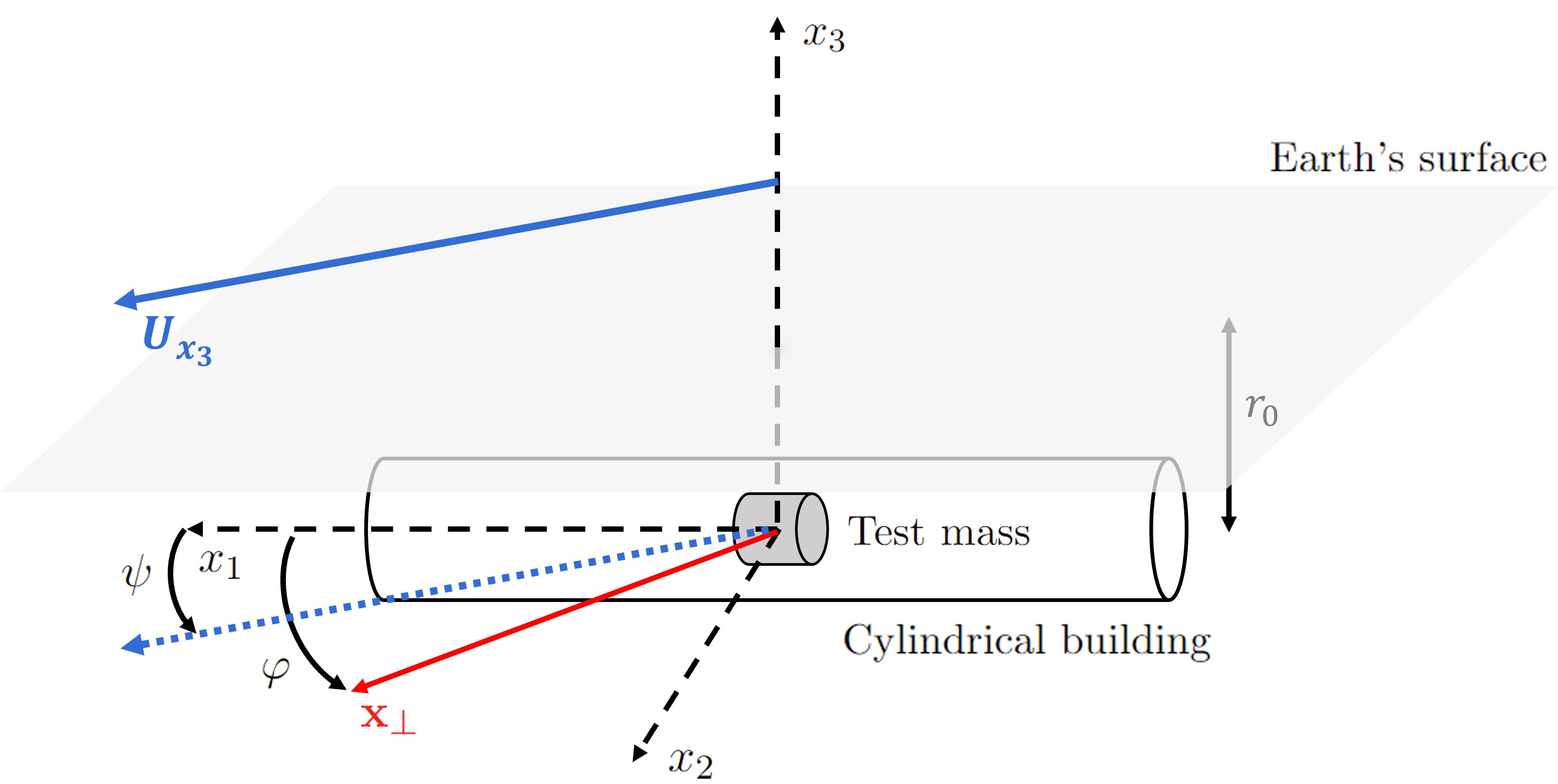}
\caption
{
Schematic representation of the geometry of the system adopted to compute the Green functions. The system of cylindrical coordinates $(x_\perp, \varphi,x_3)$ is highlighted, while $\bf U_{x_3}$ refers to the direction of the wind, parallel to the earth's surface. The dotted blue arrow is the projection of the wind onto the plane of the test mass, while $\psi$ is the angle between the detector arm, which is taken along $x_1$, and the wind direction.
}
\label{FigGeometry}
\end{center}
\end{figure}

Direct calculation of the  acceleration fluctuation from the temperature field using \cref{strain} is hopeless. The best we can do is to characterize the gravity gradient noise with its spectral density (power spectrum) $S_g(\omega,r_0)$
\begin{equation}\begin{split}
S_g(\omega, r_0)=&
\int\d^3x\d^3x'\, G_\k(\x,r_0)G_\k(\x',r_0)\CT_\omega(\x,\x')
\label{S_g}
\end{split}
\end{equation}
where

\beq
\CT_\omega(\x,\x')&=\int\d t\, \CT(\x,\x';t)\ex^{\im\omega t},
\nonumber
\\
&\equiv
\int\d t\, 
\langle\tilde T(\x,t)\tilde T(\x',0)\rangle\ex^{\im\omega t},
\eeq
and $G_\k(\x,r_0)$ is the spatial  Green function.
In the previous equations $\langle\tilde T(\x,t)\tilde T(\x',0)\rangle$ is the autocorrelation of the temperature  fluctuation field at two different points and at two different times. The stationarity  of the random process implies this autocorrelation to be a function of the time difference. The specific form of the Green function $G_\k(\x,r_0)$, which we report in \cref{AppGF}, depends on both the geometry of the problem and the coordinate choice. The geometry of the system and the set of coordinates used to perform the calculations are sketched in \cref{FigGeometry}. 

\section{Turbulence in the planetary boundary layer}
\label{sectturbboundlay}
Atmospheric turbulence is concentrated in the lowest portion of the troposphere, a region called the planetary boundary layer (PBL) \cite{stull1988introduction}. The structure of the PBL is strongly dependent on the orography, the wind and weather conditions, and the hour in the day. 

Let us consider the structure of the PBL on a typical sunny day---which means that the layer is unstably stratified---in the presence of strong wind. Under these conditions, the atmospheric turbulence contribution to NN is expected to be the maximum. For simplicity we consider the case of a horizontally uniform PBL developing over a plain region, and assume stationarity over the time-scales of interest.

\begin{figure}
\centering
\includegraphics[width=8cm]{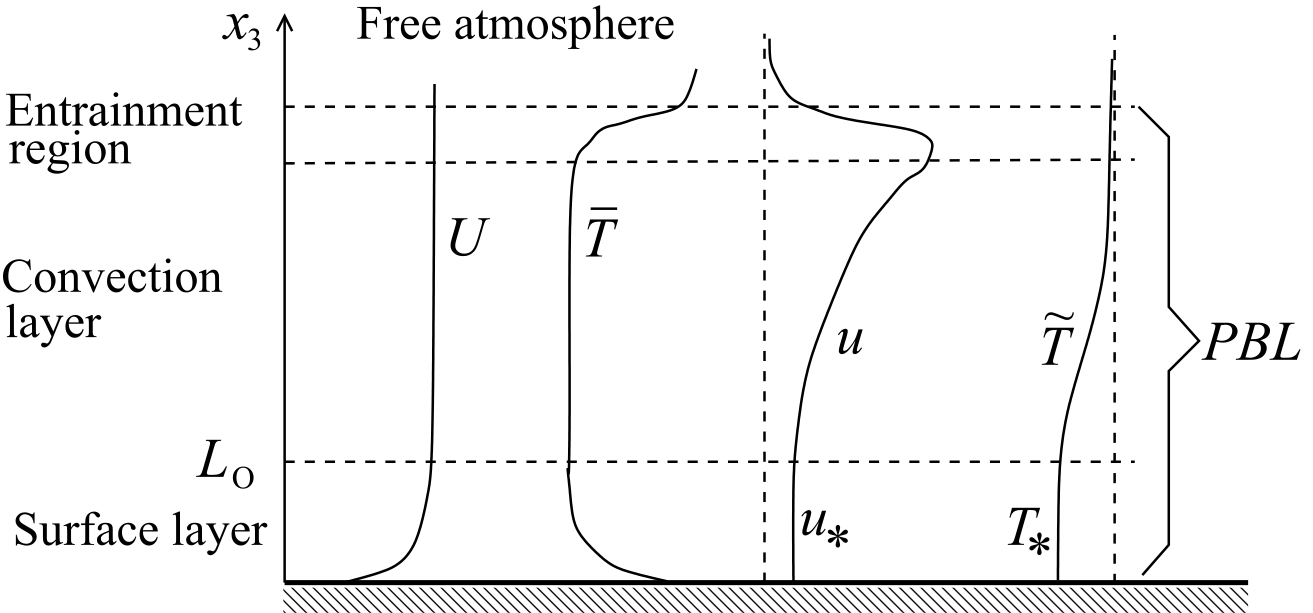}
\caption
{
Portrait (not on scale) of the PBL structure.
Vertical dashed lines indicate the zeros of the curves for turbulent velocity fluctuations $u$ and temperature fluctuations $\tilde T$.
}
\label{PBL}
\end{figure}

The PBL can be subdivided into a surface layer in which
mechanical stresses, originating from the wind interaction with the earth's surface, dominate the
dynamics, and a convection layer in which stratification is dominant. A sketch of the PBL structure is shown in \cref{PBL}.
In the surface layer, the mean wind speed $U=U_{x_3}$ 
is characterized by a logarithmic vertical profile \cite{schlichting2003boundary}
\beq
U_{x_3}\simeq \frac{u_*}{\kappa}\ln (x_3/z_0),
\label{U}
\eeq
where $\kappa=0.4$ is a universal constant called the von Karman constant, $u_*$ is the so-called friction velocity, which gives the velocity scale of the profile, and $z_0$ is an experimental parameter giving the dependence of the profile
on the roughness of the terrain. In the case of a smooth surface $z_0\simeq 0.1\nu/u_*$, where $\nu$ is the viscosity of the fluid. For \cref{U} to be valid, it is necessary that
$x_3\gg z_0$. 
Values of the roughness length $z_0$ for different types of terrain are listed in Table \ref{table1}. 
\begin{table}[]
    \centering
    \begin{tabular}{ll}
        \toprule[1pt]
        $z_0$ (m) \hspace{0.2cm}   & Terrain surface characteristics
        \\
        \midrule[0.6pt]
        1.0   & city
        \\
        0.8   & forest
        \\
        0.2   & bushes
        \\
        0.05   & farmland (open appearance)
        \\
        0.008  & mown grass
        \\
        0.005  & bare soil (smooth)
        \\ 
        0.0003  & sand surfaces (smooth)
        \\
        \bottomrule[0.8pt]
    \end{tabular}
    \caption{\label{table1}
    Schematics of terrain types and $z_0$ values \cite{troen1989european}.
    }
\end{table}

The friction velocity $u_*$ gives the amplitude of the  turbulent velocity fluctuations $u$ in the
surface layer. A similar parameter $T_*$ can be introduced,
giving the amplitude of the temperature fluctuations.
Both $u_*$ and $T_*$ are constant in the surface layer.
To characterize the strength of the wind, we shall use  the mean wind speed at a reference
height $x^{\text{ref}}_3=10\,{\rm m}$: $U_{\text{ref}}\equiv U_{10{\rm m}}$.

At height $x_3$, turbulent eddies are expected to be at most $\sim x_3$ in size, although they tend to be more elongated in the direction of the mean flow than in the vertical and spanwise directions. Similar considerations are valid for temperature fluctuations.
This identifies a characteristic time-scale of fluctuations in the surface layer
\beq
\tau(x_3)\sim x_3/u_*.
\label{tau_x3}
\eeq
We can compare the time-scale $\tau(x_3)$ with the time-scale of convection
$\tau_{\text{conv}}(x_3)\sim [\bar T x_3/(gT_*)]^{1/2}$, where $g= 9.81\, {\rm m/s^2}$ 
is the gravitational acceleration. The transition from the surface layer to the convection-dominated part of the PBL takes place for $\tau({x_3}) \sim \tau_{\text{conv}}(x_3)$, which defines (minus) the Obukhov length \cite{obukhov1971turbulence}
\beq
\LO=\frac{\bar Tu_*^2}{\kappa gT_*}.
\label{Obukhov}
\eeq
For $\bar T=300\,{\rm K}$, $T_*=1\,{\rm K}$, $U_{\text{ref}}=20\,{\rm m/s}$ and $z_0=0.05\,$m, corresponding to $u_*\simeq 1.5\,{\rm m/s}$, we get $\LO\simeq 170\,{\rm m}$.

For $x_3>\LO$ the mean velocity and temperature profiles become almost constant, and, in the presence of unstable stratification, turbulence consists of thermal plumes whose dynamics is determined by the convection time $\tau_{\text{conv}}(\LO)$.

We can determine the amplitude of the velocity and temperature fluctuations in the convective layer by requiring that, in stationary conditions, the turbulent heat flux is independent of the height, $u_{x_3}\tilde T_{x_3}\sim{\rm constant}$, where  $u_{x_3}$ and $\tilde T_{x_3}$ are the typical velocity and temperature fluctuations at height $x_3$. From $u_{x_3}\sim x_3/\tau_{\text{conv}}(\LO)$, we then get, for $x_3>\LO$,
\beq
u_{x_3}\sim \frac{u_*x_3}{\LO},
\qquad
\tilde T_{x_3}\sim \frac{T_*\LO}{x_3}.
\label{convective}
\eeq
Thermal plumes accelerate as they rise in the convective layer,
and temperature gradients concurrently diminish. The rise of the thermal plumes stops at the top of 
the convective layer, at a typical height $x_3=1\div 2\,{\rm km}$,  in the entrainment layer,
where stratification becomes strongly stable.

\subsection{Turbulence microstructure}
\label{sec:TurbMicrostr}

Turbulent structures at scale $\LO$ or higher are expected to generate a contribution to the noise spectrum at frequencies much below those of interest for GW detectors. The most significant contribution is likely to come from fluctuations close to the earth's surface, and from 
small turbulent structures originating---through the Kolmogorov cascade---from larger eddies higher in the boundary layer. The statistical properties of turbulence can be quantified in terms of correlation functions for  velocity fluctuations $C(\x,\x';t)$ and those for temperature fluctuations $C^T(\x,\x';t)$ appearing in \cref{S_g}. In HI turbulence, the spatial dependence of $C(\x,\x';t)$   $C^T(\x,\x';t)$ is  on $\x-\x'$  only, so that we have $C(\x,\x';t)=\CHI(\x-\x',t)$, $C^T(\x,\x';t)=\CHIT(\x-\x',t)$. Hence, turbulent fluctuations can be described in terms of one-time wavevector spectra 
$\CHI_k=\int\d^3x\, \CHI(\x,0)\ex^{-\im\k\cdot\x}$ and
$\CHIT_k=\int\d^3x\, \CHIT(\x,0)\ex^{-\im\k\cdot\x}$. At sufficiently small scales, in the so-called inertial range, 
turbulent fluctuations obey the Kolmogorov scaling \cite{kolmogorov1991local}
\beq
\CHI_k&\sim\flux^{2/3} k^{-11/3},
\label{C^HI}
\\
\CHIT_k&\sim\frac{\flux_T}{\flux^{1/3}} k^{-11/3}.
\label{C^HI_T}
\eeq
The dimensional constants $\flux$ and $\flux_T$ in \cref{C^HI,C^HI_T} give the speed with which turbulent fluctuations at a given scale are converted to smaller-scale turbulent fluctuations. Viscous forces become dominant at sufficiently small scales, and this determines the size $\eta_{\rm K}$ of the smallest eddies (Kolmogorov scale):
\beq
\eta_{\rm K} \sim\flux^{-1/4} \nu^{3/4},
\label{Kolmogorov scale}
\eeq
where $\nu\simeq 0.15\,{\rm cm^2/s}$ is the kinematic viscosity of air. If $L$ is the characteristic size of the largest eddies in the flow (the so-called integral scale of vortices), and $u_{\rm L}$ and $\tilde T_{\rm L}$ are their characteristic velocity and temperature scales respectively, 
\beq
\flux\sim u_{\rm L}^3/L,
\qquad
\flux_T\sim u_{\rm L}\tilde T_{\rm L}^2/L,
\label{fluxes HI}
\eeq
and \cref{C^HI,C^HI_T} will apply provided that $k L\gg 1$. 
For $u_L\sim u_*=1.5\,{\rm m/s}$ and $0.1\,{\rm m}<L<\LO=170\,{\rm m}$, we would have $0.02\,{\rm m^2/s^3}<\flux<3.4\,{\rm m^2/s^3}$, corresponding to values of the Kolmogorov length $\eta_{\rm K}$ below the millimeter.

To any given scale $l$ we can associate a characteristic velocity scale 
$\Delta_l u\sim |{\bf u}(\x+{\bf l})-{\bf u}(\x)|\sim (k^3C_k^{\text{HI}})^{-1/2}_{k\sim l^{-1}}$. From \cref{C^HI}, then, we can define an eddy turnover time 
\beq
\tau_k\sim  \left(\frac{l}{\Delta_l u}\right)_{l=k^{-1}} \sim  \flux^{-1/3}k^{-2/3},
\label{tau_k}
\eeq
which gives the typical lifetime  of turbulent structures of size $k^{-1}$.

One may expect that at height $x_3$ in the PBL, turbulent microstructure at scale $k^{-1}\ll x_3$ could be considered, using HI turbulence concepts, as originating from eddies of size $L\sim x_3$ and velocity scale $u_{x_3}$. 

Indeed, things are more complicated, at least in the surface layer, because structures with $kx_3\gg 1$ could be part of cascades originating from structures of size $L\sim x'_3$ at height $x'_3>x_3$.
The problem is less serious in the convective layer, where thermal plumes continuously carry smaller eddies upwards and prevent the simultaneous presence of multiple cascades at any given height.
We shall deal with this issue by representing turbulence as a superposition of contributions from horizontal layers of different thickness, in which turbulence is approximated as homogeneous and isotropic.

\section{Newtonian noise from homogeneous isotropic turbulence}
\label{sectNNhom}

The simplest model of NN production by atmospheric temperature fluctuations we can devise is realized by treating the PBL as an infinite layer of HI turbulence. 
The correlation function  $\CT(\x,\x';t)$ has a rather intricate structure in which  space and time dependence  are intertwined. Performing a successful analysis of $\CT(\x,\x';t)$ heavily relies on the possibility of introducing some kind of  factorization between its space and time dependence. 
The space dependence of the correlation function $\CT(\x,\x';t)$ 
in \cref{S_g} is determined by the spectrum $\CHIT_k$ in \cref{C^HI_T}. The time structure of $\CT(\x,\x';t)$ is more complex. \Cref{tau_k}
tells us that there is a multiplicity of decay times associated with the different
spatial scales in the problem. The situation is complicated by the so-called Taylor sweep, 
which consists of  turbulent structures being transported by both the mean wind and larger turbulent
structures, while being stretched and deformed into smaller-scale turbulent structures.

The analysis of the problem is facilitated by the ordering $U>u$ which
allows us, in the first approximation, to neglect the contribution of larger eddies to the transport of eddies at any given scale. A similar approach has been used in the design of some turbulent closures \cite{belinicher1987scale}. In this approximation, it is possible to eliminate
the effect of the Taylor sweep by shifting to the reference frame of the mean
wind,
\beq
\langle\tilde T(\x,t)\tilde T(0,0)\rangle=
\langle\tilde T(\x-\U t,t)\tilde T(0,0)\rangle_\smalU,
\label{U ref frame}
\eeq
where $\langle\tilde T(\x,t)\tilde T(0,0)\rangle_\smalU$ can be expressed, using \cref{C^HI_T}, as a superposition
of Fourier modes, each decaying at the time-scale fixed by \cref{tau_k}:
\beq
\langle\tilde T(\x,t)&\tilde T(0,0)\rangle_\smalU
=\frac{\flux_T}{\flux^{2/3}}
\nonumber
\\
&\times
\int\frac{\d^3k}{(2\pi)^3}\frac{\d\omega}{2\pi}
k^{-11/3}h(\tau_k\omega)\ex^{\im(\k\cdot\x-\omega t)}.
\label{TTh}
\eeq
We assume the function $h$ to be symmetric, normalized to 1, and going to zero
for large values of the argument. 
Equation (\ref{TTh}) tells us that, in Fourier space, the correlations for temperature fluctuations can
be expressed as the product of $h(\tau_k\omega)$ times a function of $k$. We substitute now \cref{U ref frame,TTh} into \cref{S_g}, and obtain after simple algebra,
\beq
S_g
=\frac{\flux_T}{\flux^{2/3}}\int&\frac{\d^3k}{(2\pi)^3}\, k^{-13/3}
|G_\k(r_0)|^2
\nonumber
\\
&\times h\left[\tau_k(\omega-\k\cdot\U)\right].
\label{int}
\eeq
The Green function $G_\k(r_0)$ has been computed in \cref{AppGF}. For the case of HI turbulence, it reads (see \cref{G_k HI})
\beq
|G_\k(r_0)|^2=
\left(\frac{2\pi\alpha\cos\phi}{k}\right)^2\ex^{-2k_\perp r_0},
\label{G^2 HI}
\eeq
where, we recall,
$\k_\perp\equiv (k_1,k_2,0)$, $\cos\phi=k_1/k_\perp$, $r_0$ is the depth of the detector, and $\alpha= G \bar \rho/\bar T$ is the conversion factor from temperature to acceleration fluctuations  [see \cref{strain}].
In \cref{sectresults} we  will use \cref{int}, together with the Green function \eqref{G^2 HI}, as the starting point for the numerical computation of the spectral density for NN generated by HI turbulence.

We note the small-$k$ divergence of the integrand in \cref{int}, which requires the function $h$ to decay sufficiently fast for large values of the argument.

The fast decay of the frequency spectrum $h$ reflects the character of the time decorrelation process as the result of the continuous stretching and scrambling of turbulent structures by the turbulent flow. This is to be opposed to the case of thermal fluctuations, where a microscopic mechanism (molecular motion) is at play, causing the correlation function $C(t)$ to be not differentiable at $t=0$, and the associated spectrum to decay like $\omega^{-2}$ at large $\omega$. We can verify that, with such a slow decay of frequency spectrum, the integral in \cref{int} would be logarithmic divergent at small $k$. 

Similar difficulties in following a Langevin-equation-based approach in turbulence have been discussed in \cite{wu2021stochastic}, concerning subgrid modeling for large-eddy simulations of wall flows.

\subsection{Frozen turbulence limit}
\label{ftl}

An important limit of \cref{int}, which has been explored in \cite{Creighton:2000gu}, is that of frozen turbulence, in which the time-decay of correlations in the reference frame of the wind is slow and the time-decorrelation of $\CHIT(\x;t)$ is only a consequence of the Doppler shift induced by the mean wind. The frozen turbulence limit is realized by approximating the function $h$ in \cref{int} with a Dirac delta, which requires  the width of $h$, seen as a function of $k$, to be much smaller than both $U/\omega$ and the width of $|G_\k(r_0)|^2$. Inspection of \cref{int,G^2 HI} gives us the condition
\beq
\tau_{\omega/U}\gg\max(r_0/U,\omega^{-1}),
\label{cond1}
\eeq
which means that the structures contributing to the noise must have an eddy turnover time
significantly longer than both the inverse of the frequency and the transit time over
a distance $r_0$. The second condition is especially interesting: $r_0/U$ represents
the effective time the turbulent structure is effectively seen by the detector, and \cref{cond1} tells us that, for a frozen turbulence hypothesis to be satisfied, the depth at which the detector is situated must not be too large. 
In frozen turbulence conditions, \cref{int} takes the form 
\beq
S_g^{ft}
&=\frac{\flux_T}{\flux^{1/3}}\int\frac{\d^3k}{(2\pi)^2}\, k^{-11/3}
|G_\k(r_0)|^2\, \delta(\omega-\k\cdot\U).
\label{frozen}
\eeq
Combining \cref{G^2 HI,frozen} produces the general expression
\beq
S^{ft}_g=\hat S_g^{ft}(\psi, \omega r_0/U)\frac{\alpha^2\flux_TU^{8/3}}{\flux^{1/3}\omega^{11/3}},
\label{frozen dimensional}
\eeq
where $\hat S_g^{ft}$ is dimensionless, and  $\psi$ is the angle between $\U$ and the detector arm (see \cref{FigGeometry}). 
The ratio $\flux_T/\flux^{1/3}$ in the formula coincides
with the parameter $c_T^2$ in \cite{Creighton:2000gu}, giving $\flux_T/\flux^{1/3}=0.2 \ {\rm K}^2 \ {\rm m}^{-2/3}$. 

The integral in \cref{frozen} can be computed analytically. The details of the calculations are  described in  \cref{app:analSHI}, and the final result is a combination of hypergeometric functions (see \cref{hyper}).

The small frequency regime of the spectrum can be directly inferred from \cref{frozen dimensional}. As discussed in \cref{AppHI}, $\hat S_g^{ft}$ has as a finite limit $S_g^{ft}(\psi,0)=S_g^{ft}(\psi)$, which implies a  power-law scaling in $\omega$ for the dimensional spectrum $S_g^{ft}$ (this can also be verified by taking the limit of the exact expression \eqref{hyper}). 

The regime $\omega r_0/U\gg 1$ is analyzed
in \cref{AppHI} and gives us the exponential behavior
\beq
S^{ft}_g\sim \frac{\flux_T\alpha^2 U^{8/3}\cos^2\psi}{\flux^{1/3}\omega^{11/3}}
\left(\frac{U}{\omega r_0}\right)^\frac{1}{2}
\exp\left(-\frac{2\omega r_0}{U}\right).
\label{frozen ld}
\eeq
It is quite interesting to compare our \cref{frozen ld} with the results of Ref.~\cite{Creighton:2000gu}, where also the NN generated by HI turbulence in the frozen limit has been analyzed. We see that the exponential behavior of the noise spectrum as a function of $\omega$  and $r_0$ found in Ref.~\cite{Creighton:2000gu} holds only for frequencies $\omega$ much higher than $U/r_0$. 

\subsection{Weak wind regime}
The weak wind regime is realized for $\omega\tau_{\omega/U}\ll 1$, which 
corresponds to approximating 
$h[\tau_k(\omega-\k\cdot\U)]\simeq h(\omega\tau_k)$ in \cref{int}.
The analysis in \cref{AppHI} produces, in this case, the power-law behaviors
\beq
S^{ww}_g\sim\frac{\flux_T\alpha^2}{\flux^{2/3}}\times
\begin{cases}
k_\omega^{-16/3} r_0^{-2},& k_\omega r_0\gg 1,
\\
k_\omega^{-10/3},& k_\omega r_0\ll 1,
\end{cases}
\label{weak wind}
\eeq
where
\beq
k_\omega=\flux^{-1/2}\omega^{3/2}
\label{komega}
\eeq
is the inverse size of vortices with eddy turnover time $\omega^{-1}$.

\section{Newtonian noise from turbulence in the PBL}
\label{sec:WallTurbulence}

In this section,  we go beyond the isotropic and homogeneous approximation  for turbulence by building a more realistic model, in which the vertical structure of the PBL is fully taken into account. Conversely, we continue to assume homogeneous correlations horizontally. Following an approach described in \cite{naguib1992investigation}, we model turbulence in the PBL as a superposition of HI turbulence contributions in horizontal layers $[0,\bar x_3]$, where $\bar x_3$ ranges from a minimum height $x_{3, \rm min}\sim z_0$---marking the transition to the region where turbulence is strongly affected by the roughness geometry---to a maximum height $L_\PBL$ identifying the top of the PBL. As a rule of thumb, one usually sets $x_{3, \rm min}=10z_0$ \cite{schlichting2003boundary}. The latter choice, however, could cut potentially important contributions from vortices at the lower end of the logarithmic region. In our model, therefore, we choose a value in-between, ${x_{3, \rm min}}= \ex z_0$,\footnote{The choice is arbitrary, but it allows a smooth transition, in the vertical wind speed profile, from a linear behavior in the roughness layer, $U_{x_3} \propto x_3/x_{3, \rm min}$, to the logarithmic profile above $U_{x_3}\propto \ln\left(\ex x_3/x_{3, \rm min} \right)$.} to identify the smallest integral-scale temperature fluctuations contributing to the noise. These temperature fluctuations have amplitude $T_*$ and eddy turnover time $\tau_{x_{3, \rm min}^{-1}}\sim x_{3, \rm min}/u_*$. Thus, a smaller $x_{3, \rm min}$ will correspond to a stronger high-frequency contribution to the noise. Indeed, we will show in \cref{sectresults} that the specific choice of $x_{3, \rm min}$ in terms of $z_0$ only alters the high-frequency portion of the spectral density, which is however always below the sensitivity curve of ET for reasonable values of the parameters (see bottom panel of \cref{fig:wall_varying_xmin}). 

The contributions to the temperature fluctuation $\tilde T(\x,t)$ from layers with $\bar x_3\ge x_3$, $\tilde T(\x,t|\bar x_3)$, are assumed uncorrelated. The spectrum in each layer can be assumed to obey Kolmogorov scaling only for $k\bar x_3\gg 1$. We must thus extend \cref{C^HI_T,tau_k} to integral scales $k\bar x_3\le 1$. The simplest possibility is to assume a sharp transition into the Kolmogorov scaling regime exactly at $k\bar x_3= 1$, i.e. to set
\beq
\CHIT_k(\bar x_3)&\sim\frac{\tilde T_{\bar x_3}\bar x_3^3}{\max\left(1,
(k\bar x_3)^{11/3}\right)};
\label{CHIT PBL}
\\
\tau_k(\bar x_3)&\sim\frac{\bar x_3}{u_{\bar x_3}
\max\left(1,(k\bar x_3)^{2/3}\right)},
\label{tau_k PBL}
\eeq
where \cref{fluxes HI} has been used.

For simplicity, the effect of the Taylor sweep in each
layer is approximated with that of a constant wind $U_{\bar x_3}$. The temperature correlation resulting from the superposition of the contributions in Eqs. (\ref{CHIT PBL}) is in the form
\beq
\CT_{\k_\perp,\omega}(x_3,x_3')
&=\int_{\max(x_3,x'_3)}^{L_\PBL}\frac{\d\bar x_3}{\bar x_3}
\int \frac{\d^3 k}{2\pi}
\CHIT_k(\bar x_3)\nonumber\\
&\times
h[\tau_k(\bar x_3)(\omega-\k\cdot\U_{\bar x_3})]\ex^{\im k_3\cdot(x_3-x_3')}.
\label{long mess}
\eeq
As in the case of \cref{int}, the spectrum $h$ is evaluated at the Doppler-shifted frequency $\omega-\k\cdot\U_{\bar x_3}$, generated by expressing correlations in the reference frame of the wind in the layer $\bar x_3$. Note the factor $\bar x_3^{-1}$ in the integral, which guarantees that all layers have equal weight.

We substitute \cref{long mess} together with \cref{CHIT PBL,tau_k PBL} and the propagator \eqref{G_k PBL} calculated in \cref{AppGF} into \cref{S_g} and obtain, after straightforward algebra,
\beq
S_g&=\frac{\alpha^2}{\pi}\int_{x_{3, \rm min}}^{L_\PBL}\d\bar x_3\int_0^{+\infty}\d k
\int_0^1 \d p \,\frac{p}{\sqrt{1-p^2}}\nonumber\\
&\ \ \times\frac{\bar x_3^3\tilde T^2_{\bar x_3}A(\k,\bar x_3)B(\k,\bar x_3)\ex^{-2k p r_0}}
{U_{\bar x_3}\max[1,(k\bar x_3)^{13/3}]},
\label{S_g PBL}
\eeq
where $p=k_\perp/k$, and
\beq
A&=\int_0^{2\pi}\d\phi\,\cos^2\phi \ h[\tau_k(\bar x_3)(\omega-\k\cdot\U_{\bar x_3})],
\label{A}
\\
B&=k^2\int_0^{\bar x_3}\d x_3\d x'_3\,\ex^{\im k_3(x_3-x_3')-k_\perp(x_3+x_3')}
\nonumber
\\
&=1+\ex^{-2k_\perp x_3}-2\ex^{-k_\perp x_3}\cos(k_3\bar x_3).
\label{B}
\eeq

In \cref{sectresults} we will use \cref{S_g PBL} as a starting point for the numerical computation of the spectral density for NN generated by turbulence in the PBL.
We note the scaling $\tilde T^2_{\bar x_3}/U_{\bar x_3}\propto \bar x_3^{-3}$ for $\bar x_3>\LO$
in \cref{S_g PBL}, which descends from \cref{convective}. This suggests that the dominant contribution to $S_g$ comes from the surface layer, where $\tilde T^2_{\bar x_3}/U_{\bar x_3}=T_*^2/u_*$. Numerical analysis confirms this indication. This prompts us to replace $L_\PBL \to \LO$ in \cref{S_g PBL}, thus approximating the noise production in the PBL with the contribution in the surface layer.

To analyze the contribution to the noise in the surface layer, it is convenient to shift to the so-called wall units \cite{schlichting2003boundary}, which we identify with a hat:
\beq
\hat\omega=\frac{\omega z_0}{u_*},\quad \hat r_0=\frac{r_0}{z_0}, \quad \hat L_{\rm O}=\frac{\LO}{z_0}.
\label{wall units}
\eeq
Note from Table \ref{table1}, that for reasonable values of $r_0$ and $z_0$, we have always $\hat r_0\gg 1$. In terms of wall units, \cref{S_g PBL} can be written in the form
\beq
S_g=\frac{\alpha^2T_*^2z_0^3}{u_*}\hat S_g(\hat\omega,\hat r_0,\hat L_{\rm O}),
\label{S_g hat}
\eeq
with $\hat S_g$ dimensionless.

\subsection{Limit behaviors}
\label{subsec Limit behaviors}
The asymptotic analysis in Appendix \ref{asympt PBL} 
tells us that, in the two limits of small and large 
$\hat\omega$, the dominant contribution to $\hat S_g$ comes from turbulent structures in the integral range ($k\bar x_3<1$) and in the inertial range 
($k\bar x_3>1$), respectively. This is associated with the observation that the 
large $\omega$ contribution to $S_g$ is increasingly concentrated at
smaller $\bar x_3$.

The mechanism for the separation between small and large frequency behavior can be understood by considering that the frequency $\omega$ selects, as dominant contributors to $S_g$, turbulent structures with size $k_\omega^{-1}$, such that either the eddy turnover time $\tau_k$ or the transit time $(kU_{\bar x_3})^{-1}$ is $\sim\omega^{-1}$. A maximum frequency for the production of integral range vortices of size $k_\omega^{-1}$ is then identified by $k_\omega z_0\sim 1$, as smaller integral-scale vortices would have to reside inside the roughness of the terrain.

The different scaling of the contributions to $S_g$ from integral and inertial range turbulent structures is associated with a transition from a small to a large frequency behavior that is characterized by an increase in the decay rate with frequency. Indeed, the asymptotic analysis in \cref{asympt PBL} gives us, for $\omega^{-1}$ shorter than the eddy turnover time of integral scale vortices at height $z_0$,
\beq
\hat S_g\sim \hat\omega^{-8}\hat r_0^{-2},\quad \min(\hat\omega,\hat r_0)\gg 1,
\label{large omega}
\eeq
which is \cref{large omega 1}. This equation tells us that we will always have a $1/r_0^2$ behavior of $S_g$ at large frequencies and for large values of $r_0$.

For small $\hat\omega$, instead, we have two possible behaviors depending on the magnitude of $\hat r_0$ (see \cref{smallish omega}):
\beq
\hat S_g\sim
\begin{cases}
\hat x_{3, \rm {max}}^3, 
& 1\gg \hat\omega\hat r_0
\\
\hat r_0^{-2} \hat x_{3,\rm {max}}^5,
&\hat\omega\hat r_0\gg 1\gg\hat\omega,
\end{cases}
\label{small omega}
\eeq
where
\beq
\hat x_{3,\rm{max}}=\min(\hat\omega^{-1},\hat L_{\rm O}),
\label{x3max}
\eeq
and where the ratio $r_0/u_*$ in $\hat\omega\hat r_0=\omega r_0/u_*$ is the eddy turnover time of integral scale vortices at height $r_0$. Thus, at small frequencies, power-law decay in $r_0$ of the noise spectrum will ensue for $r_0\sim u_*/\omega$, which is the size (and the height) of integral scale vortices with eddy turnover time $\omega^{-1}$. 
As already remarked, for reasonable values of $r_0$ and $z_0$, we always have $\hat r_0\gg 1$, which is why we have chosen to disregard the range $\hat\omega\gg 1\gg\hat r_0$ in \cref{large omega} in the first place. Notice, moreover, that the $1/r_0^2$ behavior arises also in the weak-wind regime of the HI approximation (see \cref{weak wind}). As we show in  \cref{sec:Scalingr0}, this is a general behavior arising whenever the effect of eddy decay dominates over wind advection and $r_0$ is large.  

\section{Results}
\label{sectresults}

In this section we numerically compute and analyze the behavior of the power spectra obtained from \cref{S_g} for the three different cases discussed in this paper: $(a)$ HI turbulence in the frozen limit, $(b)$ HI turbulence with finite correlation time, and $(c)$ turbulence in a horizontally homogeneous PBL. Specifically, for case $(a)$, we use the expression \eqref{frozen}, while for cases $(b)$ and $(c)$, as starting point for the numerical computation of the power spectra, we use the more general \cref{int,S_g PBL}, respectively. The extended set of data presented and related  to this work  are available online \cite{GITHub_repo}. Moreover, we compare the resulting spectra with the ET-D sensitivity curve, to investigate the possible impact of atmospheric NN on next-generation GW detector measurements. Note that the ET-D sensitivity curve is expressed in terms of the strain power spectrum $S_h$, which is related to the acceleration power spectrum $S_g$ in \cref{S_g} by the relation

\beq
S_h=4\frac{S_g}{\omega^4L_\text{arm}^2},
\label{rescale}
\eeq
where $L_{\rm arm}$ is the length of the detector arm, while the factor of $4$ takes into account the number of test-masses in the detector. Indeed, one can assume that the NN contributions on each test mass are uncorrelated, since those arising from correlated signals between two or more test masses are expected to be negligible \cite{Creighton:2000gu,Har2019}. We  rescale the modeled spectra according to \cref{rescale}. \\

\noindent\textbf{(a) HI turbulence: frozen limit}---In the frozen limit, the frequency spectrum of the temperature fluctuations (function $h$ in \cref{int}), is simply a Dirac delta (see \cref{frozen}). This allows us to use the analytic expression for the integral \eqref{frozen} in terms of a sum of hypergeometric functions provided in \cref{hyper}.
\begin{figure}
    \centering
    \includegraphics[scale=0.3]{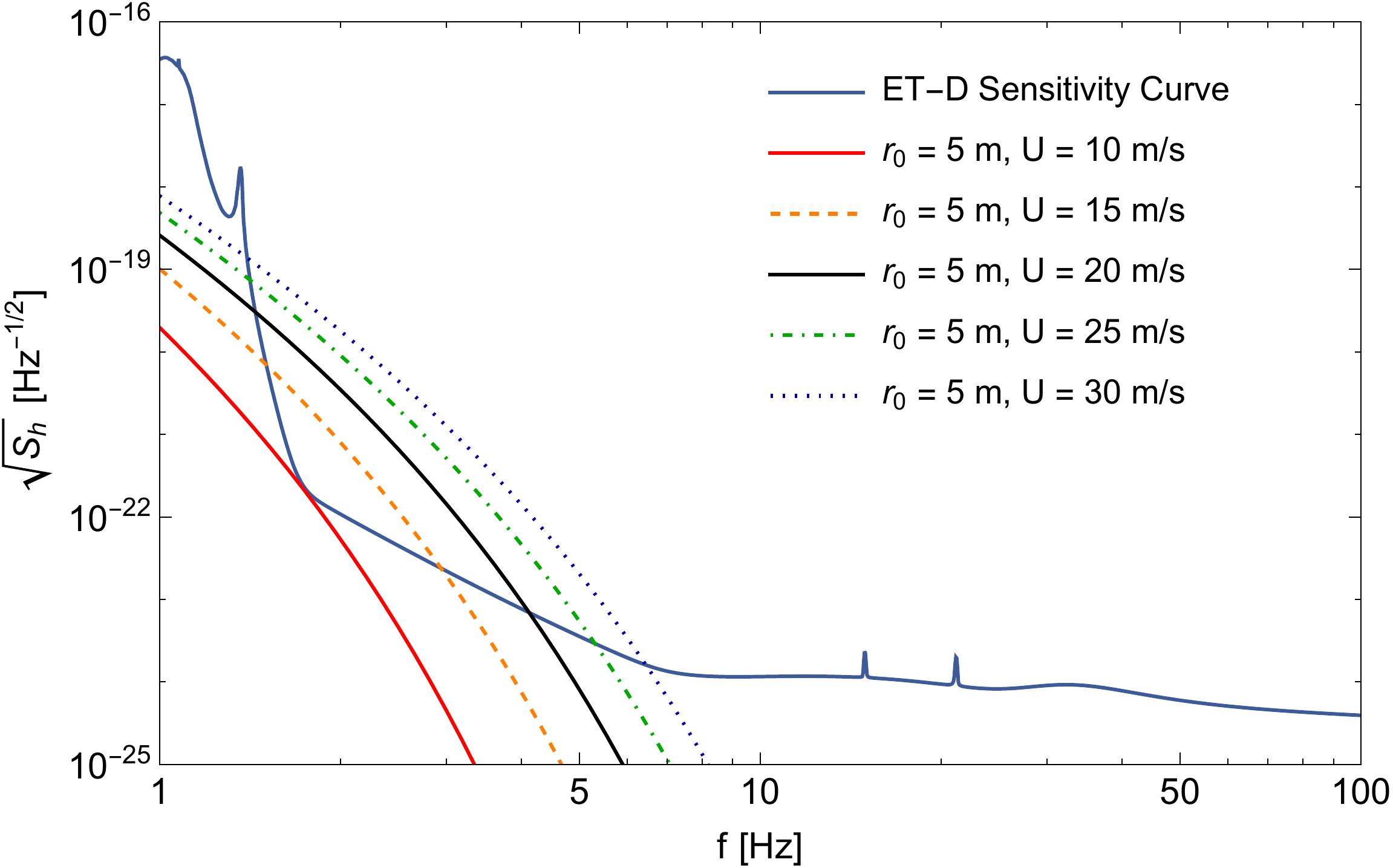}
    \includegraphics[scale=0.3]{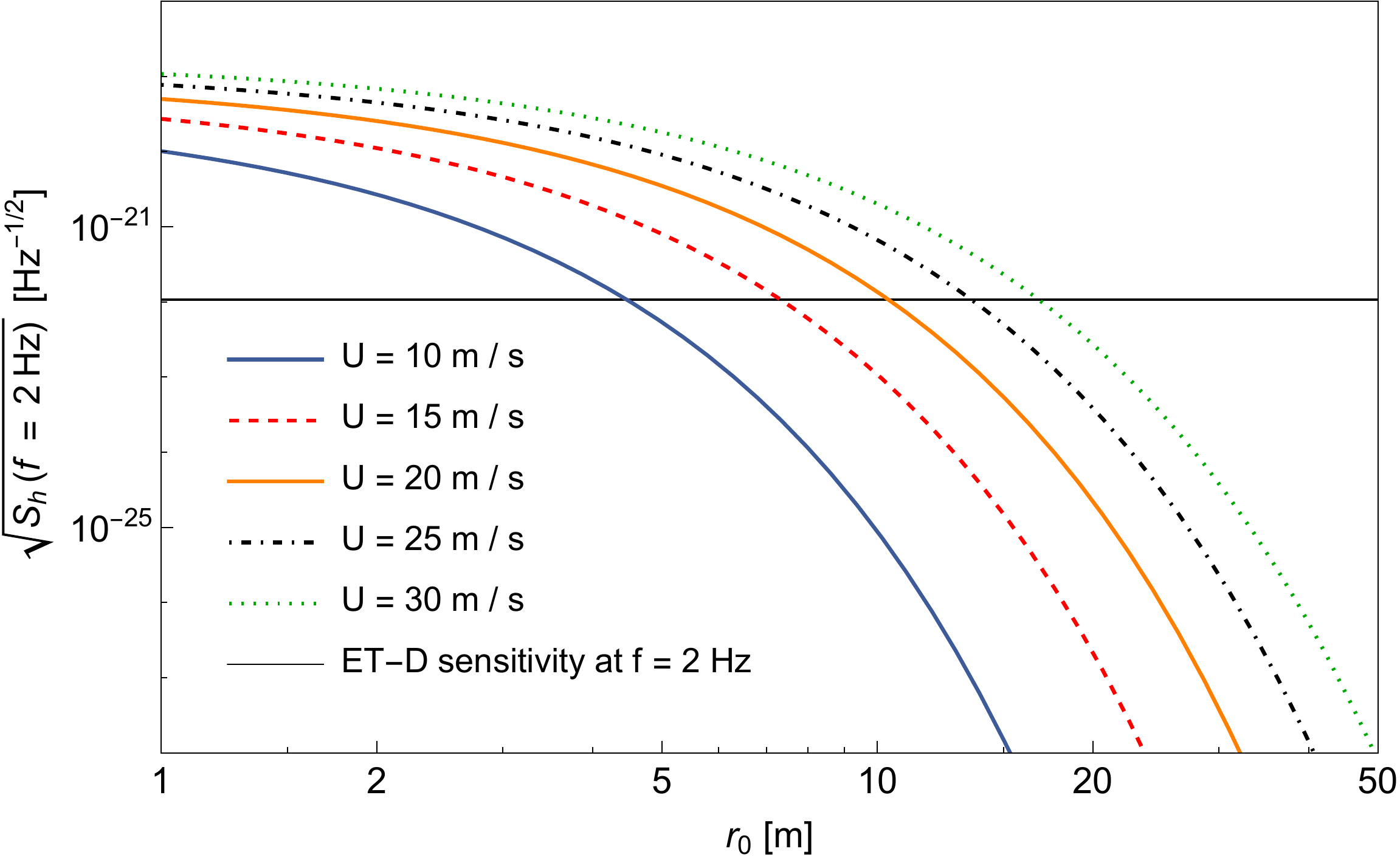}
    \caption{
    HI turbulence - frozen limit. {\bf Top panel}: noise spectra as functions of the frequency $f=\omega/(2\pi)$ at fixed detector depth $r_0$ and with varying wind speed $U$. {\bf Bottom panel}: strain spectrum for $f = 2 \text{ Hz}$ as a function of the detector depth $r_0$ for selected values of the wind speed $U$. In all cases we have taken $\psi=0$.
    }
    \label{fig:noise_HI_ftl}
\end{figure}
The noise curves obtained in this limit, for fixed $r_0$ and varying $U$, are shown in the upper panel of \cref{fig:noise_HI_ftl} and compared with the sensitivity curve of ET-D. We also show the variation of the strain spectrum at the fixed pivotal value of the  frequency of $2$ Hz, as a function of the depth $r_0$ (bottom panel of \cref{fig:noise_HI_ftl}). For all the curves in \cref{fig:noise_HI_ftl}, we have set the coefficient $c_T^2=\flux_T/\flux^{1/3}$ in \cref{frozen} equal to its value  in \cite{Creighton:2000gu}, 
$c_T^2 = 0.2\text{ K}^2 \ \text{m}^{-2/3}$. Here, and in the following cases (HI turbulence with finite correlation time, and inhomogeneous turbulence), we have set $\psi=0$, which corresponds to wind blowing parallel to the detector arm (see subsection (c) for a discussion of the dependence of the spectra on $\psi$).

As already discussed below \cref{frozen ld}, the NN is exponentially suppressed for both large frequencies and large detector depths. In particular, it is worth noting that the noise curve is below the ET sensitivity curve for $r_0\gtrsim 30 \text{ m}$ and for frequencies $f\gtrsim 5 \text{ Hz}$, where $f = \omega / (2\pi)$. Moreover, the larger the wind speed, the larger the noise amplitude, as explicitly shown in \cref{fig:noise_HI_ftl}. The bottom panel of \cref{fig:noise_HI_ftl} also shows that, for $f= 2 \ \rm Hz$, if $r_0\lesssim5\text{ m}$ and  
$U\ge 10\,{\rm m/s}$, the NN spectrum is comparable or above the sensitivity of ET-D.  The same happens if $r_0 \lesssim 20 \ \rm m$ and $U\simeq 30 \ \rm m/\rm s$. The frequency range considered in the plot is not wide enough to visualize the power-law behavior predicted by \cref{hyper} in the $\omega \ll U/r_0$ limit. We have verified that  the latter arises at very small frequencies $f\ll 1\text{ Hz}$, for the values of $U$ and $r_0$ considered here. Alternatively, very large values $U$ or very small values of $r_0$ would be required. If the  interferometer is built underground, however, $r_0$ is at least $\sim 1 \ {\rm m}$; on the other hand, winds stronger than $30 \ \text{m/s}$ are very unlikely.\\

\noindent\textbf{(b) HI turbulence: finite correlation time}---To go beyond the frozen turbulence approximation, an explicit form of the function $h$ in \cref{int} must be selected. However, such a choice is not straightforward. In fact, the functional dependence of $h$ on the physical parameters relies on the underlying dynamics of the turbulent structures, whose detailed knowledge is still missing. A reasonable and likewise simple form for the temporal correlation spectrum in \cref{int}, which satisfies the conditions given in \cref{sectNNhom}, is a Gaussian function
\beq
h[\tau_k(\omega-\mathbf{k}\cdot\mathbf{U})]=\ex^{-\tau_k^2(\omega-k_\perp U \cos(\phi-\psi))^2}.
\label{hgauss}
\eeq
We have verified that other choices of $h$, such as that of a top-hat function, do not produce significantly different results. Using \cref{hgauss}, the integral~\eqref{int} has no closed-form solution, thus the noise spectra must be computed via numerical integration.

The integrand in \cref{int} as well as in \cref{S_g PBL}  is concentrated in a tiny section of the integration domain, whose shape is highly dependent on the parameters involved. This makes numerical integration by quadratures cumbersome. The integrals have thus been computed  using the VEGAS algorithm \cite{Vegas}, which exploits a Monte Carlo technique with importance sampling and is ideally suited for multidimensional integrals. 
It should be pointed out that, with this approach, the numerical estimation is highly computational consuming ($\sim3$ core-hour for a single value of $S_h$). Therefore, the integrals have been computed by using the HYDRA framework \cite{hydra}, which is designed to perform data analysis and numerical integration tasks on massively parallel platforms, commonly used in the high-energy, particle physics community. Using this tool, we carried out the integrals in \cref{int,S_g PBL} obtaining our results with a sub-percent precision in a reasonable amount of time.

We have repeated the analysis in \cref{fig:noise_HI_ftl}, for different values of $U$ and $r_0$, and of the new parameter $\flux$, which sets the scale of the eddy turnover time $\tau_k$ (see \cref{tau_k}). We have considered values of $\flux$ in the range $0.01 \  \text{m}^2/\text{s}^3\lesssim\flux\lesssim 1 \text{ m}^2/\text{s}^3$ (see discussion following \cref{fluxes HI}). For consistency with the frozen turbulence case, we have taken $\flux_T=c_T^2\flux^{1/3}$, with $c_T^2=0.2\,{\rm K^2m^{-2/3}}$. As in  the previous subsection, we limited our analysis to the case $\psi=0$.
\begin{figure}
    \centering
    \includegraphics[scale=0.3]{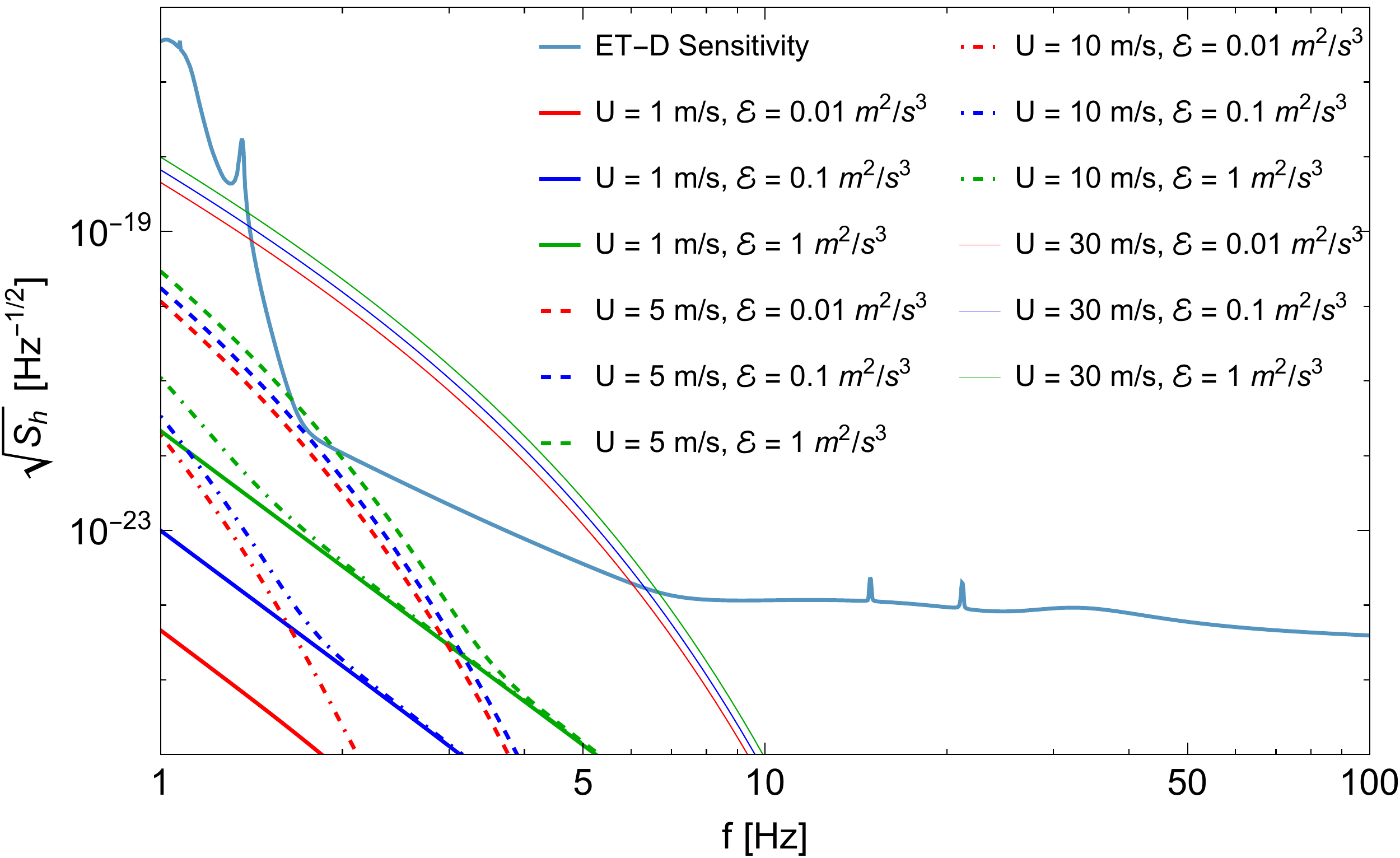}
    \includegraphics[scale=0.3]{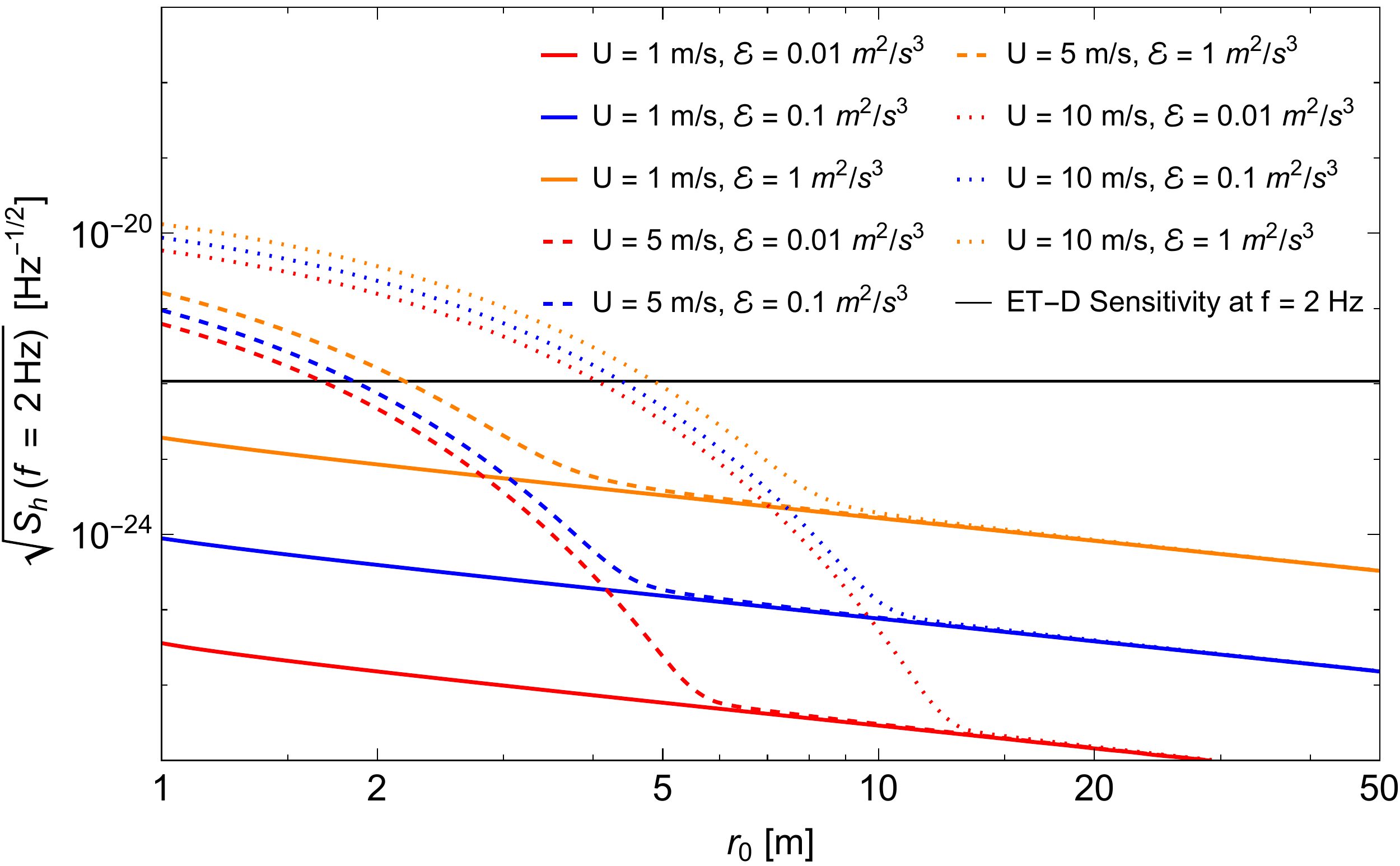}
    \caption{
    HI turbulence - finite correlation time. {\bf Top panel}: noise spectra as a function of the frequency $f$ for $r_0 = 5\text{ m}$ and $\psi=0$ and for selected values of the parameters in the HI approximation with Gaussian time correlations. {\bf Bottom panel}: noise spectra as a function of the detector depth $r_0$ for $f = 2\text{ Hz}$ and $\psi = 0$ and for selected values of the parameters in the HI approximation with Gaussian time correlations.
    }
    \label{fig:HI_decay}
\end{figure}

The results are shown in \cref{fig:HI_decay}. The curves are quite similar to those in \cref{fig:noise_HI_ftl}. In general, the noise still decays for large detector depths and large frequencies, while it increases with the wind speed and with $\flux$. The latter reflects the increase in the amplitude of the temperature fluctuations with $\flux_T=c_T^2\flux^{1/3}$, which implies, through \cref{tau_k}, an increase of the characteristic frequency of the fluctuations as well. However, a new feature, not present in the frozen regime, is observed in both panels, namely a transition to a power-law regime in $f$ and $r_0$ if $f$ is very large, $U$ is very small or $r_0$ very large
(see \cref{weak wind}). At intermediate values of the parameters (take, for instance, the curve with $U=5 \ \rm m/\rm s,\, \flux= 0.1 \ \rm m^2/\rm s^3$ in the top panel of \cref{fig:HI_decay}), the  noise curves show a transition from exponential (at low frequencies) to power-law (at large frequencies) behavior. As expected, the noise curves show a weak dependence on $\flux$ as long as the dominant contribution to the noise comes from wind transport rather than from vortex decay. This occurs for $U\gtrsim 10 \ {\rm m/s}$. This feature makes the noise curves for HI turbulence with Gaussian time correlation and those obtained in the frozen approximation comparable in the parameter region of interest, i.e that in which the noise curves are close to the ET-D sensitivity curve.
Indeed, it can be seen that, in this exponentially-damped regime, the NN  from  turbulence with finite correlation time and frozen turbulence have roughly the same impact on  the GW detector. On the contrary, the frequency at which the transition between the two regimes (exponential damping and power-law scaling) occurs, strongly depends on the value of $\flux$. In particular, we see that the power-law behavior shifts to larger frequencies for smaller $\flux$.

\begin{figure}
    \centering
    \includegraphics[scale=0.3]{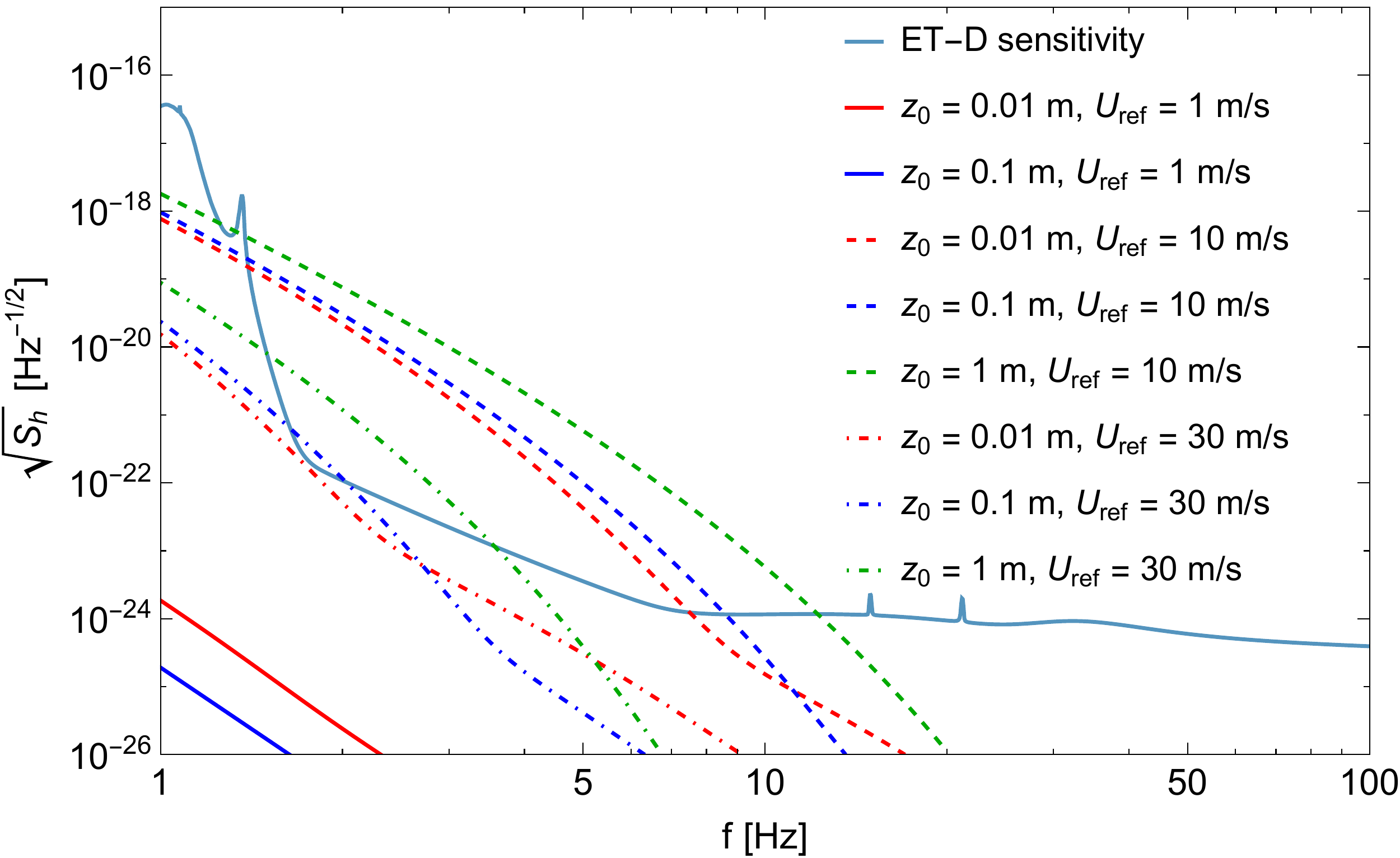}
    \includegraphics[scale=0.3]{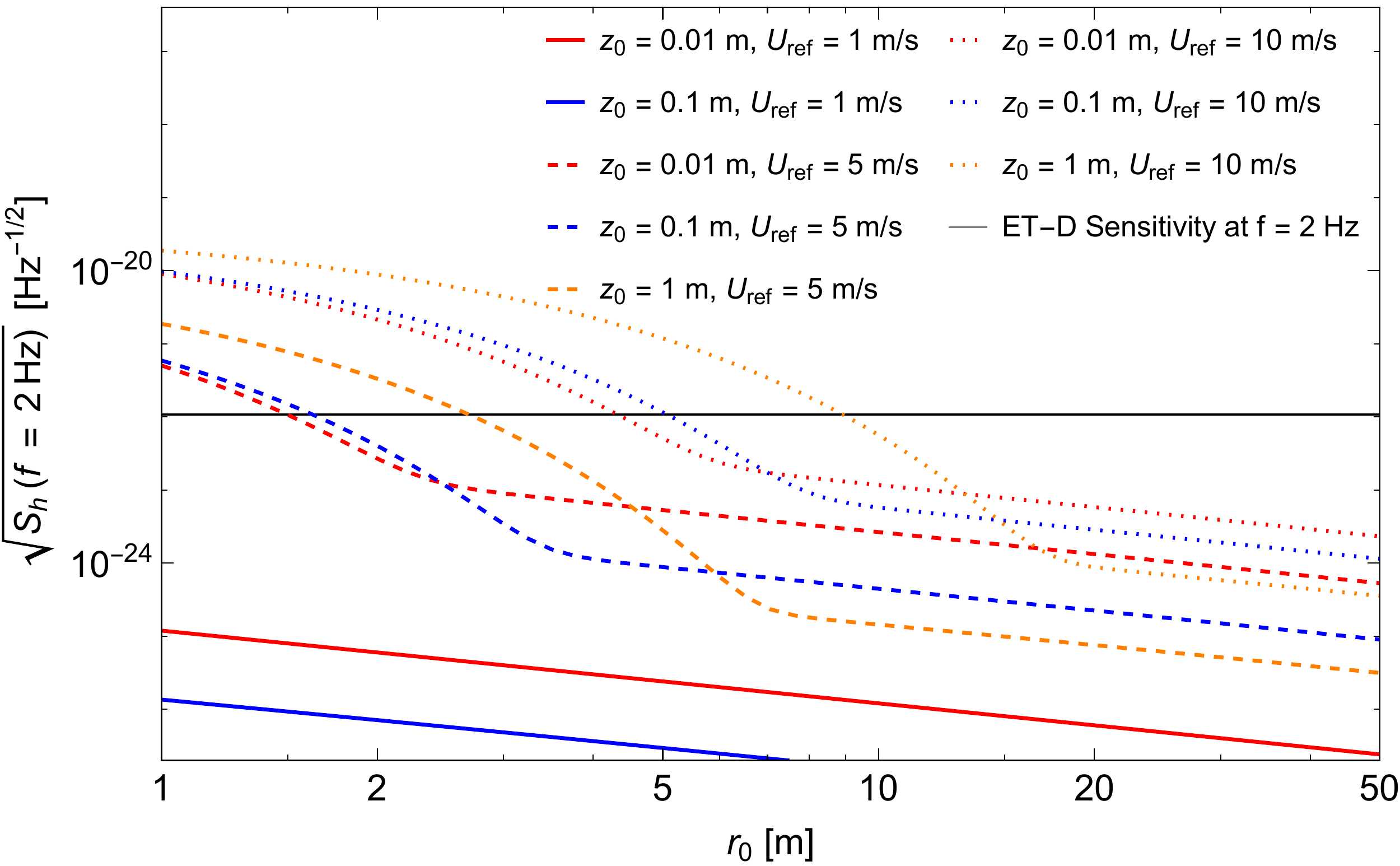}
    \caption{Inhomogeneous turbulence. \textbf{Top panel}: noise spectra as a function of the frequency $f$ for $r_0 = 5\text{ m}$ and $\psi = 0$ and for selected values of the parameters in the surface-layer model with Gaussian time correlations. \textbf{Bottom panel}: noise spectra as a function of the detector depth $r_0$ for $f = 2 \text{ Hz}$ and $\psi = 0$ and for selected values of the parameters in the surface-layer model with Gaussian time correlations.}
    \label{fig:wall}
\end{figure}

Inspection of \cref{fig:HI_decay} also shows that going beyond the frozen approximation changes only slightly the impact of both the wind speed $U$ and the depth $r_0$ on power spectrum noise curves at small frequencies. The differences between the noise curves for HI turbulence with Gaussian time correlation and those for frozen turbulence are quite small. At a fixed wind speed of $10\  \rm m/\rm s$, for instance, both  curves cuts the ET-D sensitivity curve at $r_0\approx 4-5 \ \rm m$.\\

\noindent\textbf{(c) Inhomogeneous turbulence}---The noise spectrum is now described by \cref{S_g PBL}. 
We have proceeded as in the (b) case, and assumed Gaussian time correlations, as described in \cref{hgauss}. The noise spectrum has been evaluated from \cref{S_g PBL}, as a function of the experimentally accessible quantities $U_{\rm ref}$, $z_0$ and $T_*$. The quantity $T_*$, which gives the scale of the temperature fluctuations, enters the expression for $S_g$ \eqref{S_g PBL} as a scale factor and has been fixed at the reference value $T_*=1\text{ K}$. For $z_0$ we have taken values corresponding to situations ranging from that of bare soil to that of a forest or a city district, as described in \cref{table1}. As in the previous HI turbulence cases, we have set $\psi=0$.

The noise amplitude increases with the wind speed $U_{\rm ref}$ and, for small frequencies, with the roughness length $z_0$. This is not surprising, considering that the turbulence intensity, parameterized by $u_*$, is proportional to $U_{\rm ref}$ and it increases with $z_0$ through \cref{U}. In the high-frequency region, instead, in correspondence with the onset of the power-law behavior, the noise curves are characterized by  $\sqrt{S_h}$ scaling as $z_0^{-3/2}$, as predicted by \cref{large omega,S_g hat}. We also recall that a significant contribution to high frequencies fluctuations is produced in the region near the ground, where the parameter $x_{3, \rm min}\sim z_0$ plays the role of a cutoff (see \cref{S_g PBL}); we will return to this point at the end of the section. We note that, since $T_*$ is the same in all curves, the increase in the noise amplitude with $U_{\rm ref}$ and $z_0$ is a consequence of the increase of the characteristic frequency of the fluctuations only.

The curves in the top panel of \cref{fig:wall}, like those in the top panel of \cref{fig:HI_decay}, are characterized by a transition from an exponential to a power-law behavior at sufficiently small values of $U_{\rm ref}$ and $z_0$. A transition to a power-law, this time for $r_0$, is observed also in the bottom panel of \cref{fig:wall}, analogous to the one observed in \cref{fig:HI_decay}. The crossover point in $r_0$  shifts to the right for large values of $U_{\rm ref}$ and $z_0$, which is consistent with the predictions in the asymptotic large $\omega$ and small $\omega$ limits provided by \cref{large omega,small omega}. This is not surprising, since $U_\text{ref}$ plays in the inhomogeneous model the same role played by $U$ in the HI model. 
For very small frequencies, moreover, the spectra are characterized by another power-law scaling, which however can be observed only for $f \lesssim 2\text{ Hz}$ and for large values of $U_\text{ref}$.

Another interesting feature is the scaling with the parameter $z_0$, describing the roughness of the terrain. Indeed, the spectra scale as $\sim z_0^{-3}$ for $\omega\gg u_\ast/z_0$ while, for $\omega\ll u_\ast/z_0$, they first increase with $z_0$ and then they become almost independent of this quantity (see \cref{large omega,small omega}). Similarly to what happens with the parameter $\flux$ in the HI case, the weak dependence of the noise curves on the parameter $z_0$ is a characteristic of the regime where wind transport dominates over vortex-decay. Again, this is a nice feature, allowing for easy comparison of the curves for inhomogeneous and homogeneous turbulence in the regions where they are close to the ET-D sensitivity curve.

\begin{figure}
    \centering
    \includegraphics[scale=0.34]{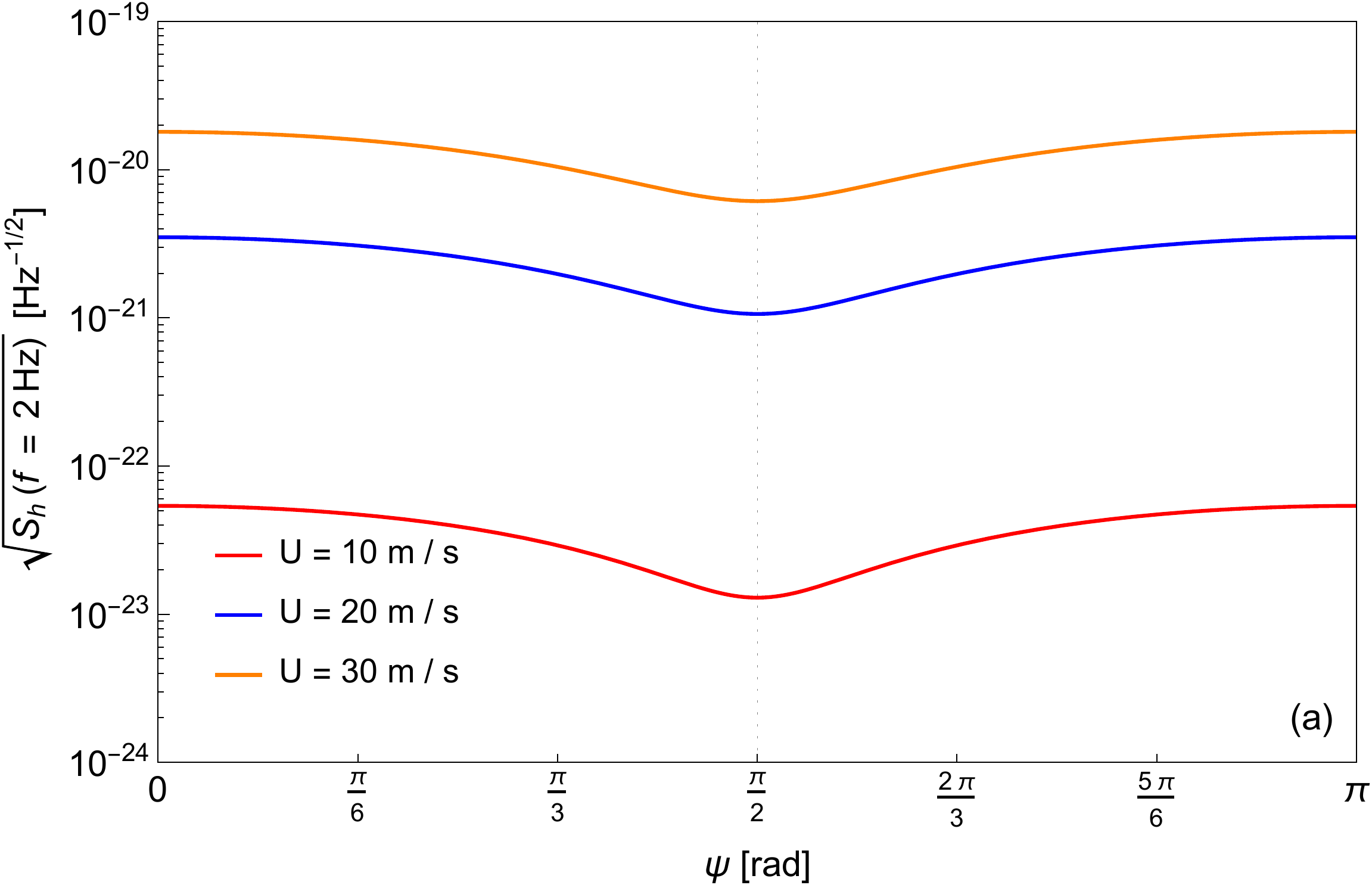}
    \includegraphics[scale=0.34]{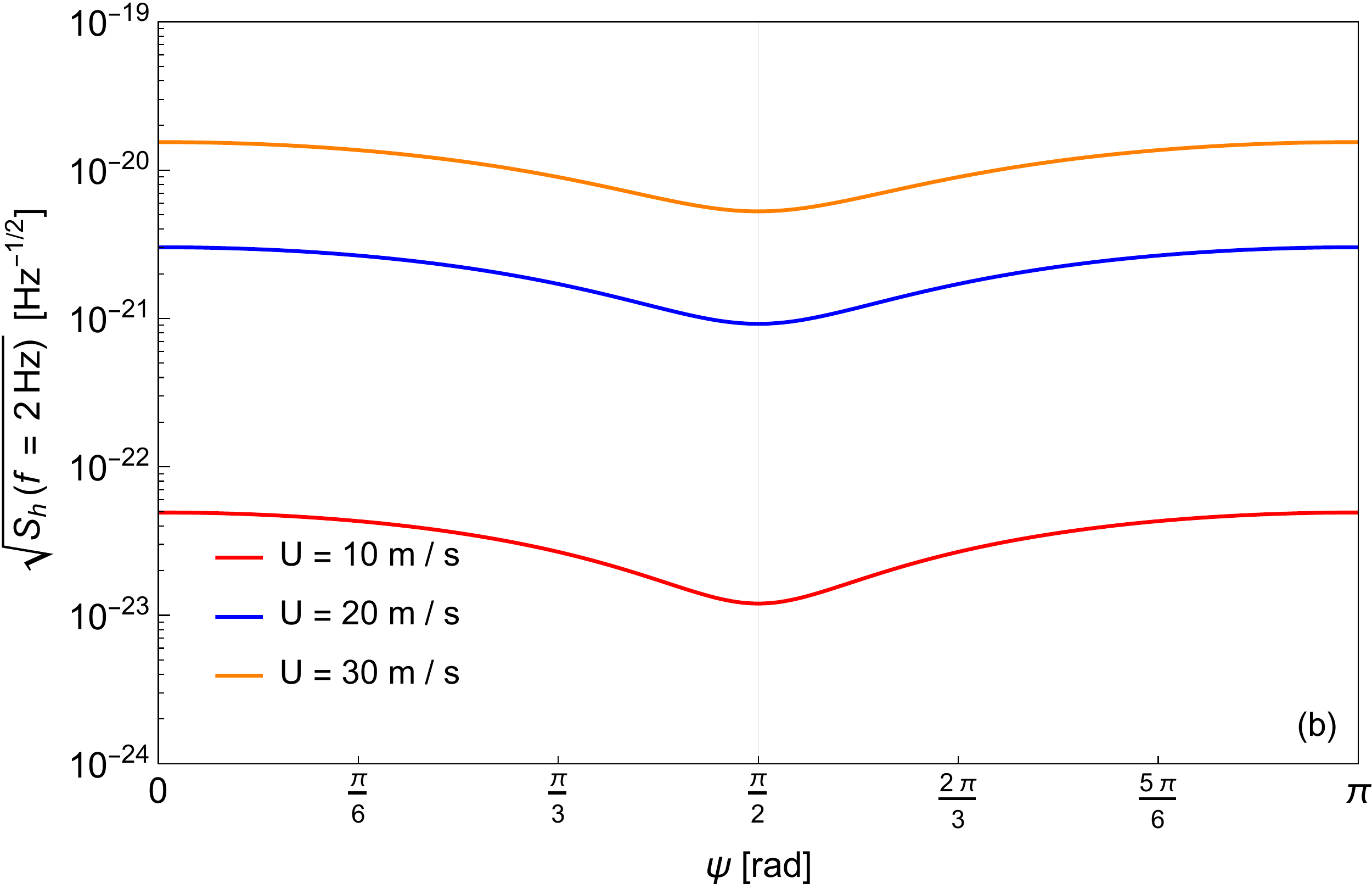}
    \includegraphics[scale=0.34]{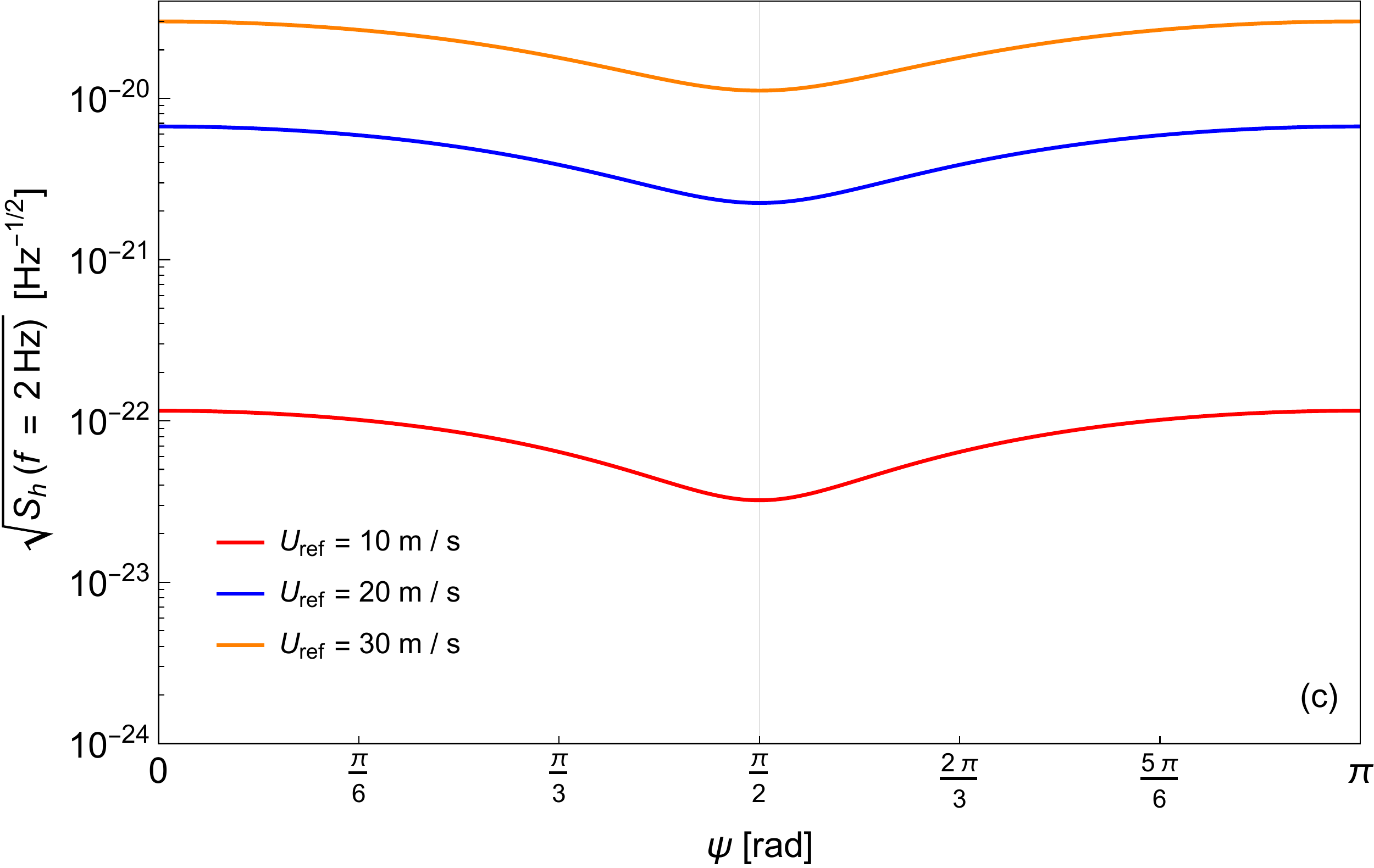}
    \caption{
    Noise spectra as a function of $\psi$ in the three cases of HI turbulence - frozen case $(a)$; HI turbulence - finite correlation time $(b)$ and inhomogeneous turbulence $(c)$, at varying $U$ (upper two panels) and $U_{\rm ref}$ (lowest panel). \\
    We fixed the values of the other parameters to: $f=2\,{\rm Hz}$, $r_0=5\,{\rm m}$, $\flux=0.1\,{\rm m^2/s^3}$, $z_0=0.1\,{\rm m}$.
    }
    \label{fig:psi}
\end{figure}

The last interesting point is the dependence of the spectra on $\psi$. Indeed, in some regimes (see, e.g., \cref{frozen ld}), the spectra could depend strongly on this parameter. Anyway, we expect $S_h$ to be the maximum for $\psi=0$. Indeed, in \cref{fig:psi}, we show that, for reasonable values of the parameters, $S_h$ varies at most by a factor $\sim 2$.

\begin{figure}
    \centering
    \includegraphics[scale=0.34]{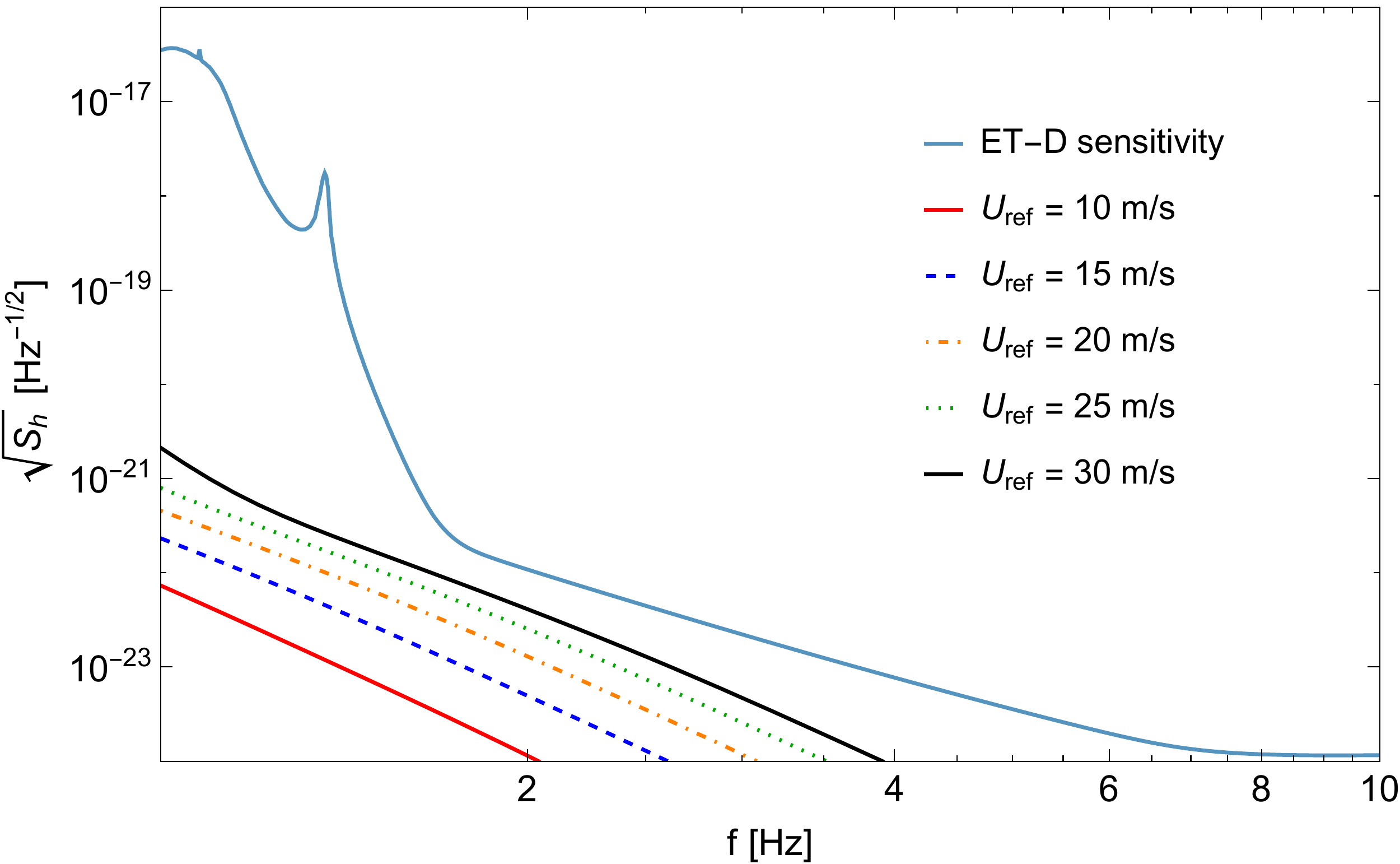}
    \includegraphics[scale=0.34]{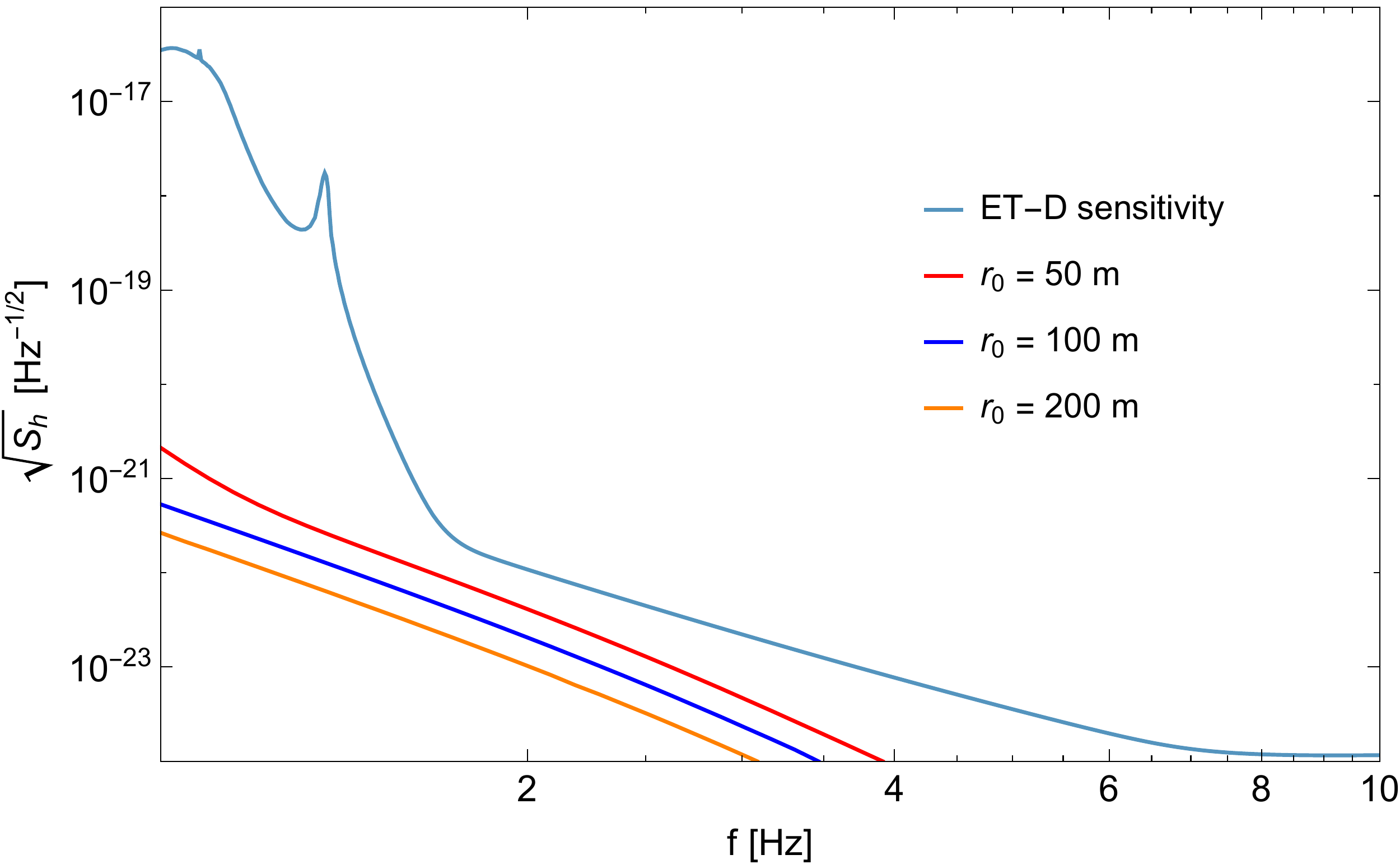}
    \caption{\textbf{Top panel}: Noise spectra for the inhomogeneous model as a function of the frequency $f$ for $r_0=50\text{ m}$ and for selected values of $U_\text{ref}$. \textbf{Bottom panel}: Noise spectra for the inhomogeneous model as functions of the frequency for selected values of $r_0$ and $U_\text{ref}=30\text{ m/s}$.\\
    We fixed the values of the other parameters to: $z_0=0.1 \text{ m}$ and $\psi=0$.}
    \label{fig:r050}
\end{figure}

On a semi-quantitative level, the results of the  inhomogeneous model confirm those obtained in the HI one (both in the general case and in the frozen approximation limit). Indeed, for fixed values of the parameters (either $U$ or $r_0$), the noise spectra in the three cases have comparable orders of magnitude, at least in the regions close to the sensitivity curve of ET-D. When the detector is located near the earth's surface, the power spectrum curves cut the ET-D sensitivity and are well above it in the frequency region $2-10 \ \rm Hz$. On the other hand, in the frequency band considered here, the noise curves go below the sensitivity curve only when the detector is located at least $50 \ \rm m$ underground (see top panel of \cref{fig:r050}). We see, however, that for large values of $U_{\rm ref}$, i.e. $U_{\rm ref}\sim 30 \ \rm m/s$, the noise curve is only a factor $\sim 5$ below the ET-D sensitivity curve,
which is a worryingly close range, considering the fact that our models are  providing only order of magnitude estimates. The situation is confirmed if we go to greater depths. Consistently with the general  scaling $r_0^{-1}$  of $\sqrt{S_h}$ derived in \cref{sec:Scalingr0}, the noise spectra are only a factor of $\sim 10-20$  below the sensitivity curve, for $r_0 \sim 100-200 \ \rm m$, respectively, (see bottom panel of \cref{fig:r050}).\\


\noindent
{\bf Wavevector cutoffs and finite-size effects}---For all the calculations in the present section, we have adopted a maximum wavevector $k_{\rm max}$ equal to the inverse of the Kolmogorov scale $\eta_{\rm K}^{-1}$ defined in \cref{Kolmogorov scale}, and we have set the parameter $x_{3,\text{min}}=\ex z_0$ as the lower bound of integration in \cref{S_g PBL}. In the inhomogeneous turbulence case of \cref{S_g PBL}, we have carried out the integral over $\bar x_3$ up to a reference height $L_{\PBL}=2\ {\rm km}$, but the contribution at $\bar x_3>L_{\rm O}$ turned out to be negligible. 

\begin{figure}
    \centering
    \includegraphics[scale=0.34]{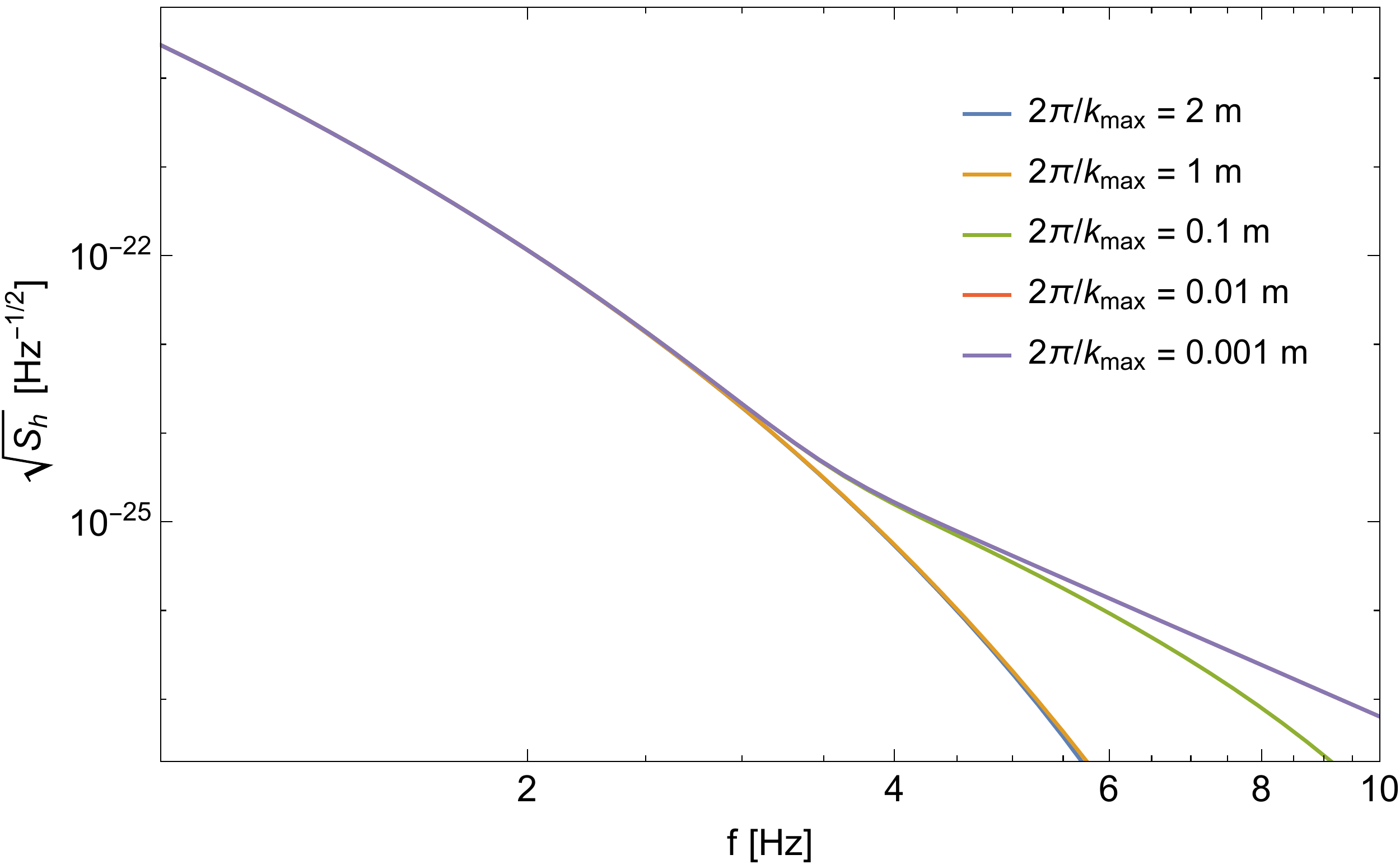}
    \includegraphics[scale=0.34]{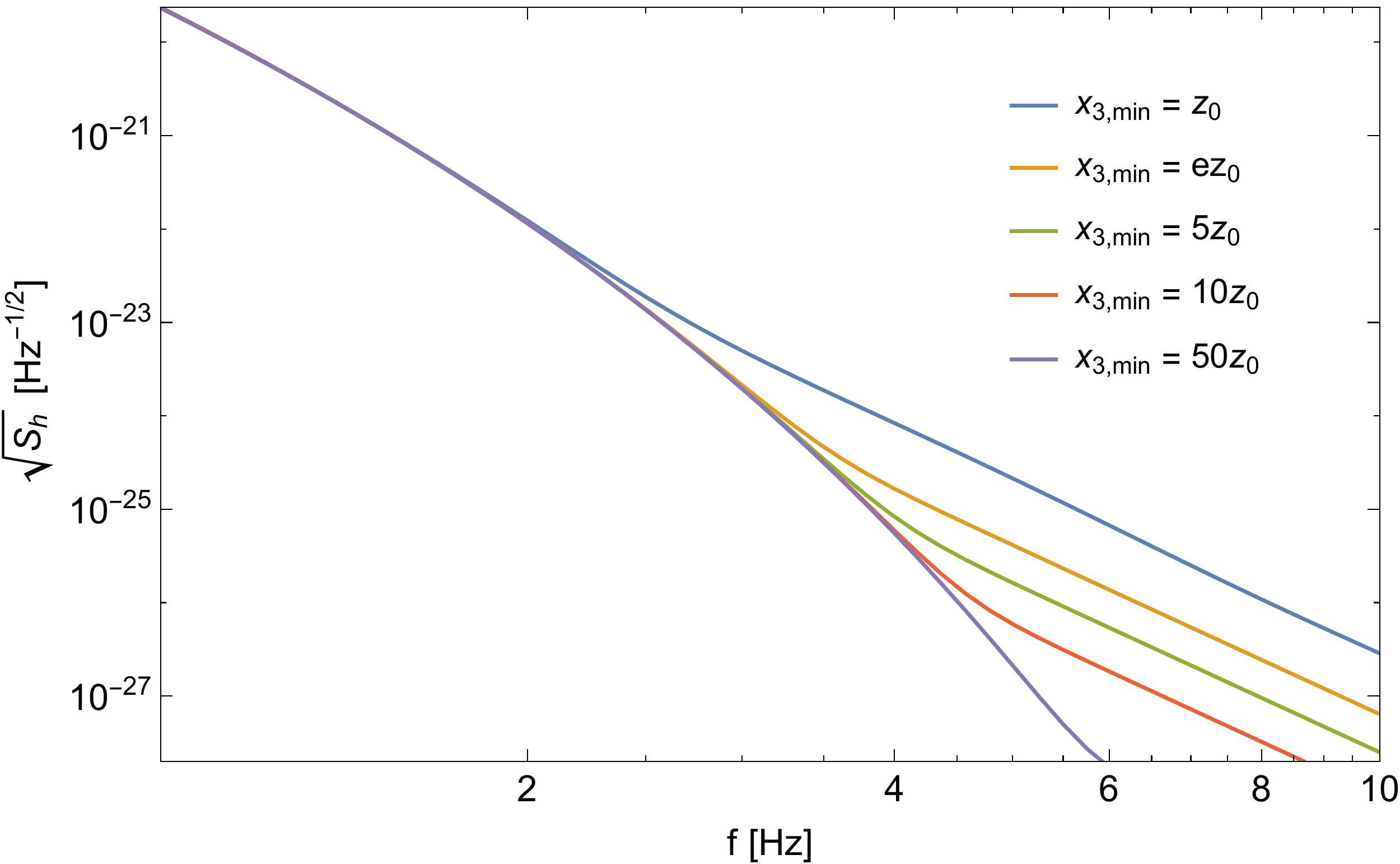}
    \caption{
    Inhomogeneous turbulence. \textbf{Top panel}: dependence of the noise spectrum on the wavevector cutoff $k_{{\rm max}}$. \textbf{Bottom panel}: dependence on the minimum height $x_{3, \rm min}$ of the vortices contributing to the noise. Values of the parameters in the two cases: $r_0=5\text{ m}$, $z_0=0.1\text{ m}$, $U_\text{ref}=10\text{ m/s}$ and $\psi = 0$.
    }
    \label{fig:wall_varying_xmin}
\end{figure}

The dependence of $S_h$ on the parameters $k_{\rm max}$ and $x_{3,\text{min}}$ is rather different. In the top panel of \cref{fig:wall_varying_xmin}, we show the effect of lowering the cutoff $k_{\rm max}$ on the noise spectrum. We limit our analysis to the inhomogeneous turbulence case,
setting, as usual, $r_0=5\,{\rm m}$ as a reference depth at which the noise is expected to be above the sensitivity threshold for the ET detector. 
We see that, when $k_{\rm max}^{-1}$ approaches the meter range, the high-frequency portion of the noise spectrum is increasingly damped. 
Inspection of the top panel of \cref{fig:wall_varying_xmin} indicates that noise-damping occurs above the kink, signaling the transition from an exponential behavior dominated by wind transport to a power-law behavior dominated by eddy decay. However, for sufficiently small $k_{\rm max}$, i.e. $2\pi k_{\rm max}^{-1}\sim 1-2 \text{ m}$, this transition does not occur in the frequency-band considered here, and only exponentially-damped ``frozen-like'' contributions are present. This is because vortices of this size have a very long decay time (see \cref{tau_k}) compared to the advection time of the wind. 

A similar situation occurs, as illustrated in the bottom panel of \cref{fig:wall_varying_xmin}, varying $x_{3,\text{min}}$.
In this case, the high-frequency portion of the noise spectrum is increasingly damped as  $x_{3,\text{min}}$ gets larger. The result is consistent with the observation in \cref{subsec Limit behaviors} that the contribution to the high-frequency portion of the NN spectrum is produced primarily by temperature fluctuations near the ground. 

The present analysis tells us that numerical simulation of the atmospheric flow, say, by large-eddy simulation \cite{stoll2020large}, would allow direct evaluation of the NN only for frequencies below a maximum that is an increasing function of $k_{\rm max}$ and $x_{3,\text{min}}^{-1}$. Frequencies above this maximum would require some kind of parameterization.

\section{Conclusions}
\label{sectconclusions}

In this paper, we have built models for the NN  generated by atmospheric turbulence, which  represent a reliable tool for the assessment of the impact of atmospheric  noise on third-generation GW detectors. 
This has been done by improving previous models for temperature-fluctuation  induced atmospheric NN. Owing to their basic assumptions (frozen in time, HI turbulence)  the latter  are not reliable enough in the frequency bandwidth and for the sensitivity levels of third generation GW detectors. This improvement has been made possible by building both models for HI turbulence with finite correlation time for temperature fluctuations and models which  also take into account  the strong inhomogeneity  of turbulence  along the vertical direction.

We have also computed the spectral density of the NN for the three classes of models as a function of the characteristic physical parameters, and compared it with the sensitivity curve of the ET detector in the xylophone configuration. The NN signal shows two kinds of regimes, one in which it behaves exponentially and the other in which it shows a power-law behavior. 
The first appears to be the signature of the dominance of wind transport and the exponential dependence on the depth of the detector, of the Green function  connecting temperature fluctuations and NN. Since in a wind-dominated regime, this dependence is weighed by a characteristic scale which  is the ratio of the wind velocity and the frequency, the result is an exponential scaling of the NN with both the depth of the detector and the frequency.
Whenever we depart from this regime (either because eddy decay becomes an important factor, or because the depth of the detector is  large), the NN dependence on the wind velocity, frequency and detector depth is a power law.
While the power spectra scale with the velocity and the frequency with a power-law which is model dependent,
 the exponent characterizing the $r_0$ scaling is fixed at $-2$ by the properties of the gravitational propagator.
 Departures from exponential behavior occur for high frequencies, or very small frequencies, the latter being out of the regime of interest for ET anyway. Moreover, in the regions of interest for ET, the NN power spectra show a very weak dependence on the parameters of the models, apart from the wind speed $U$ and the depth $r_0$. Taken together, these two facts explain why the models of HI frozen turbulence used to date worked well in the assessment of the impact of atmospheric NN for second-generation GW detectors \cite{Creighton:2000gu,Har2019}.

We have then compared the NN power spectrum calculated with our models with the sensitivity curve of ET. We have found that the atmospheric NN contribution is above the sensitivity curve in the low-frequency band when the detector is placed on the surface and/or when the wind speed is relatively large. Thus, our main result is that NN generated by atmospheric turbulence represents an important source of noise for third-generation GW detectors, which therefore must be taken into consideration and accurately analyzed, e.g. using numerical simulations. This is of paramount importance, especially if the detector has to be built on the earth's surface.

On the other hand, we have also found that passive mitigation of atmospheric NN is only partially effective. Although placing the detector underground suppresses the atmospheric NN contribution, the rather weak, $1/r_0$, decay of the noise signal implies that even an underground construction at great depth may not be enough to suppress the noise completely. Even at $r_0 \sim 200\,{\rm m}$, the noise amplitude curve, despite being always below the sensitivity  curve of ET, remains close to it in the low-frequency region. On the other hand, our modeling of turbulence provides only order of magnitude predictions.
An inaccuracy of an order of magnitude in our estimates of the power spectrum should therefore be taken into account. 
Considering this fact and the proximity of the noise amplitude to the sensitivity of the detector, at least in strong wind situations ($\sim 30\text{ m/s}$) (see \cref{fig:r050}), numerical simulations of the atmospheric flow (taking into account the orography in the detector region) and on-site measurements are advisable. 

In this regard, we expect our results to be relevant for any evaluation of NN by numerical simulation of the atmospheric flow, as atmospheric codes (such as, e.g., large-eddy simulations  \cite{stoll2020large}) have a grid scale that is typically above that of fluctuations contributing to the NN.

\section*{Acknowledgements}
We  thank the  Newtonian noise research group of the SARGRAV project and in particular Jan Harms, for helpful discussions and comments.

\appendix
\section{Green function evaluation}
\label{AppGF}

To compute the Fourier transform of the gravitational propagator, we consider the detector at depth $r_0$ below the earth's surface, which will be considered as a flat infinite plane. We first choose a Cartesian system of coordinates $x_1$, $x_2$ and $x_3$, with origin at the test mass of the detector, and with $x_1$ laying along the detector arm, while $x_3$  will be along the vertical.

Computations can be more easily performed by adopting the cylindrical system of coordinates ${\bf x}=\left(x_\perp, \, \varphi, \, x_3 \right)$, where $\varphi$ is the azimuthal angle, while $x_\perp = (x_1^2+x_2^2)^{1/2}$ refers to the direction perpendicular to $x_3$ (see \cref{FigGeometry}). Moreover, we will make use of the following expressions
\begin{subequations}
\begin{align}
&G({\bf x}; r_0)= - \alpha \ \partial_{x_1} H({\bf x}; r_0);\label{Greal}\\
&H({\bf x}; r_0)= \frac{\theta(x_3-r_0)}{\left(x_\perp^2+x_3^2 \right)^{1/2}}.\label{Hreal}
\end{align}
\end{subequations}

The Fourier transform of \cref{Hreal} then reads
\begin{equation}\begin{split}\label{Hkr0}
    H_{\bf k}(r_0) =& \int_{r_0}^{\infty} \d x_3 \int_0^{+\infty} \d x_\perp x_\perp \int_0^{2\pi}\d\varphi\\
    &\times \frac{\exp\left[-\im \left(k_3 x_3 + k_\perp x_\perp \cos\varphi \right) \right]}{\left(x_\perp^2 + x_3^2 \right)^{1/2}}
\end{split}\end{equation}
where $\k = (\k_\perp,k_3)$ and  $\k_\perp\equiv (k_1,k_2)$.

When dealing with HI turbulence, correlations have to be computed in the whole three-dimensional Fourier space. \Cref{Hkr0,Greal} together yield
\beq
G_\k(r_0)=-\frac{2\pi\alpha\cos\phi}{k_\perp+\im k_3}\ex^{-(k_\perp+\im k_3)r_0}.
\label{G_k HI}
\eeq
where we have defined $\cos\phi=k_1/k_\perp$.

In more realistic cases, things have to be treated more carefully, as we might have inhomogeneities along one or more axes. In inhomogeneous turbulence, for instance, we have inhomogeneities along the $x_3$-axis due to the inhomogeneous wind profile. Since correlations on the plane orthogonal to $x_3$ are assumed to be homogeneous in our model (see \cref{sec:WallTurbulence}), the Green function will be simply given by the $x_\perp$ and $\varphi$ integrals in \cref{Hkr0}, which yield
\beq
G_{\bf k_\perp}(x_3,r_0)=-2\pi\im \alpha\theta(x_3-r_0)\cos\phi\ \ex^{-k_\perp x_3}
\label{G_k PBL0}
\eeq\vspace{0.5cm}

Finally, since the effects of turbulence along the vertical will be integrated from the earth's surface up to infinity, we simply have to translate the origin of the system of coordinates along the $x_3$-axis by $r_0$. In other words, we simply shift $x_3 \to x_3 + r_0$, which yields
\beq
G_{k_\perp}(x_3,r_0)=-2\pi\im \alpha \cos\phi\ \ex^{-k_\perp x_3}\ex^{-k_\perp r_0}.
\label{G_k PBL}
\eeq

\section{Analytic form of the spectrum for homogeneous isotropic NN in the frozen approximation}
\label{app:analSHI}
In this appendix,  we derive  the  analytic results for the spectrum \eqref{frozen} for HI turbulence in the frozen approximation. 
We start from \cref{frozen}
\beq
S_g^{ft}=\frac{\flux_T}{\flux^{1/3}}\int\frac{\d^3\mathbf{k}}{(2\pi)^2}k^{-11/3}|G_{\mathbf{k}}(r_0)|^2\delta(\omega-\mathbf{k}\cdot\mathbf{U}).
\eeq
To compute the integral, first, we choose the geometry illustrated in \cref{FigGeometry}, and we adopt a cartesian set of coordinates in the integrated variable $\mathbf{k}$. Moreover, we choose the wind speed to be parallel to the earth's surface. This simplifies the calculations since
\beq
\mathbf{k}\cdot\mathbf{U}=k_\perp U\cos(\phi-\psi),
\eeq
where $U$
is the wind speed and $k_\perp=\sqrt{k_1^2+k_2^2}$. To get rid of the angle $\psi$ in the Dirac delta, we rotate the reference frame around the $k_3$-axis by an angle $\psi$ and we write $\cos\phi=k_1/k_\perp$. Then, using \cref{G_k HI} we get
\beq
S_g^{ft}=&\ \frac{\flux_T}{(2\pi)^2\flux^{1/3}}\int_{0}^{\infty}\d^3\mathbf{k}\,k^{-11/3}|G_\mathbf{k}|^2\delta(\omega-k_1 U)\nonumber\\
=& \ \alpha^2\frac{\flux_T}{\flux^{1/3}}\int_{-\infty}^\infty\d k_1\int_{-\infty}^\infty\d k_2\int_{-\infty}^\infty\d k_3\nonumber\\
&\nonumber\times k^{-17/3} k_\perp^{-2}(k_1^2 \cos^2\psi + k_2^2\sin^2\psi)\\
&\times\delta(\omega-k_1 U)\ex^{-2k_\perp r_0}.
\eeq

The integral over $k_3$ can be easily done and yields
\beq
\int_{-\infty}^\infty\frac{\d k_3}{k^{17/3}}&=
\sqrt{\pi}\ \frac{\Gamma\left(7/3\right)}{\Gamma\left(17/6\right)}k_\perp^{-14/3}.
\eeq
The integral over $k_1$ is instead trivial using the delta function, which simply sets $k_1 = \omega/U$.

The last integration in $k_2$ can be solved changing the integration variable, mapping $k_2\to\xi=\sqrt{1 + (U k_2/\omega)^2}$. With this substitution, the integral becomes
\begin{multline}
S_g^{ft}=2\mathcal{C}\int_1^\infty\d\xi\,\frac{\xi}{\sqrt{\xi^2-1}}\\
\ \times\frac{\cos^2\psi+(\xi^2-1)\sin^2\psi}{\xi^{20/3}}\ex^{-2\omega r_0\xi/U},
\end{multline}
where $\mathcal{C}=\alpha^2 \flux_T \sqrt{\pi} \ \Gamma(7/3)/(\flux^{1/3} \Gamma(17/6))\ U^{8/3} \omega^{-11/3}$. The integral can be done analytically and gives a combination of hypergeometric functions
\begin{widetext}
\beq
\label{hyper}
S_g^{ft}=& \  2\mathcal{C}\cos^2 \psi\,
\biggl\{\frac{\sqrt{\pi } \Gamma \left(\frac{17}{6}\right) \, _1F_2\left(-\frac{7}{3};-\frac{11}{6},\frac{1}{2};x^2\right)}{2 \Gamma \left(\frac{10}{3}\right)} + 32 x \Gamma\left(-\frac{17}{3} \right) \nonumber\\
&\times \biggl[2^{2/3} x^{14/3} \, _1F_2\left(\frac{1}{2};\frac{10}{3},\frac{23}{6};x^2\right)-\frac{26180 \sqrt{\pi } \, _1F_2\left(-\frac{11}{6};-\frac{4}{3},\frac{3}{2};x^2\right)}{6561 \ \Gamma\left(\frac{17}{6}\right)}\biggr] \biggr\}\nonumber\\
&+2\mathcal{C}\sin^2\psi\, \biggl\{\frac{\sqrt{\pi } \Gamma \left(\frac{11}{6}\right) \, _1F_2\left(-\frac{7}{3};-\frac{5}{6},\frac{1}{2};x^2\right)}{4 \Gamma \left(\frac{10}{3}\right)}+8 x \Gamma \left(-\frac{11}{3}\right) \nonumber \\
&\times \biggl[2^{2/3} x^{8/3} \, _1F_2\left(-\frac{1}{2};\frac{7}{3},\frac{17}{6};x^2\right)-\frac{55 \sqrt{\pi } \, _1F_2\left(-\frac{11}{6};-\frac{1}{3},\frac{3}{2};x^2\right)}{243 \ \Gamma \left(\frac{17}{6}\right)} \biggr]\biggr\},
\eeq
\end{widetext}
where, for convenience, we defined $x \equiv r_0 \omega/U$.

\section{Asymptotic results in homogeneous isotropic turbulence}
\label{AppHI}

\subsection{Frozen turbulence}
An explicit expression for the dimensionless spectrum $\hat S_g^{ft}$ in \cref{frozen dimensional} is obtained by substituting \cref{G_k HI} into \cref{frozen}.
Let us set $\U=(U,0,0)$ and introduce dimensionless quantities $\hat\omega=\omega r_0/U$ and $\hat k=kU/\omega$. We find
\beq
\hat S^{ft}_g=\int\frac{\d \hat k_2\d \hat k_3}{\hat k^{17/3}}\cos^2(\psi+\phi)\ex^{-2\hat k_\perp\hat\omega}
\eeq
where $\hat k_1=1$ from the Dirac delta in \cref{frozen}.
We easily verify that the integral converges to a finite function of $\psi$ for $\hat\omega\to 0$. This implies that for $\omega\ll U/r_0$, $S_g$ scales like a power in $\omega$ and $U$ (see \cref{frozen dimensional}).

In the opposite limit $r_0\gg U/\omega$, we can approximate
\beq
\hat S_g^{ft}&\simeq\int\frac{\d \hat k_2\d \hat k_3}{\hat k^{17/3}}
\left[\frac{\cos^2\psi+\hat k_2^2\sin^2\psi}{1+\hat k_2^2}\right]
\nonumber\\
&\times\exp\left[-\hat\omega(2+\hat k_2^2)\right],
\label{tmp1}
\eeq
where we have exploited $\hat k_2\sim\hat\omega^{-1/2}\ll 1$.
Evaluating the integral in \cref{tmp1} to lowest order in $\hat\omega^{-1}$ 
yields \cref{frozen ld}.

\subsection{Weak wind regime}
Let us indicate $p=k_\perp/k$. We can evaluate the integral in \cref{int} in spherical coordinates. If $h[\tau_k(\omega-\k\cdot\U)]\simeq h(\tau_k\omega)$,
the integral over $\phi$ in \cref{int} is trivial, so that we are left with
\beq
 S^{ww}_g&\sim \frac{\flux_T \alpha^2}{\flux^{1/3}}\int_0^{+\infty}\d k\,k^{-13/3}h(\tau_k\omega)
\int_0^1 \d p\, p \, \ex^{-2pkr_0}
\nonumber
\\
&=\frac{\flux_T \alpha^2}{\flux^{1/3}}\int_0^{+\infty}\d k\,\frac{1-(1+2kr_0)\ex^{-2kr_0}}{4k^{19/3}r_0^2}h(\tau_k\omega)
\nonumber
\\
&\sim \frac{\flux_T \alpha^2}{\flux^{1/3}}\int_{k_\omega}^{+\infty}\d k\,\frac{1-(1+2kr_0)\ex^{-2kr_0}}{k^{19/3}r_0^2},
\label{tmp2}
\eeq
where $k_\omega$ is defined in \cref{komega}.
We can now carry out the integral in \cref{tmp2} in the two limits
of large and small depth and we recover
\cref{weak wind}.\\

\section{Time correlations for turbulence in the surface layer}
\label{A app}

Let us indicate
\beq
\eta=\omega\tau_k(\bar x_3),
\quad
\zeta=k_\perp \tau_k(\bar x_3)U_{\bar x_3}.
\eeq
We find the following limit behaviors for the function $A$ in \cref{A}: 
\begin{itemize}
\item
If either $\zeta\ll \eta\sim 1$ or $\eta\gg 1$ and $\zeta\ll |h(\eta-\zeta)/h'(\eta-\zeta)|$,
\beq
A(\eta,\zeta)\simeq \pi h(\eta).
\label{caso1}
\eeq
\item
If either $\zeta\gg\max(\eta,1)$ or $\zeta>\eta\gg 1$,
\beq
A(\eta,\zeta)\sim \frac{\cos^2(\psi+\phi_m)}{\zeta\sin\phi_m},
\label{caso2}
\eeq
where $\phi_m=\arccos(\eta/\zeta)$.
\item
If $\eta\gg \max(1,\zeta)$ and $\zeta\gg |h(\eta-\zeta)/h'(\eta-\zeta)|$,
\beq
A(\eta,\zeta)\sim \cos^2\psi\frac{h^{3/2}(\eta-\zeta)}{|\zeta h'(\eta-\zeta)|^{1/2}}.
\label{caso3}
\eeq
\end{itemize}
The result in \cref{caso1} is straightforward. The result in \cref{caso2} is
obtained by saddle-point approximation, expanding $h(\eta-\zeta\cos\phi)\simeq h(\zeta\phi'\sin\phi_m)$, $\phi'=\phi-\phi_m$,
where $\phi_m=\arccos(\eta/\zeta)$, and then integrating from $-(\zeta\sin\phi_m)^{-1}$ to
$(\zeta\sin\phi_m)^{-1}$. The result in \cref{caso3} is obtained by
expanding
\beq
h(\eta-\zeta\cos\phi)
&\simeq h(\eta-\zeta(1-\phi^2/2))
\nonumber
\\
&\simeq h(\eta-\zeta)+(\zeta\phi^2/2)h'(\eta-\zeta)
\nonumber
\eeq
and then integrating by steepest descent. In order to have $\hat A=O(1)$, it is sufficient that $\eta\sim \zeta\sim 1$.

\section{Asymptotic results for turbulence in the surface layer}
\label{asympt PBL}

In the integral in \cref{S_g PBL} we separate contributions from integral scale vortices (``domain 1'', $k\bar x_3<1$) and those from inertial scale vortices (``domain 2'', $k\bar x_3>1$). Different cutoffs act in the integral: 
\begin{itemize}
    \item The term $\ex^{-2k p r_0}$, accounting for the decay of the signal with the depth of the
detector; 
\item 
The factor $k^{-13/3}$, associated with the decay of turbulent fluctuations
at small scales; 
\item
The function $A$ that filters eddies at time-scales below
$\omega^{-1}$.
\end{itemize}
The analysis in \cref{A app} tells us that the function $A$ is surely negligible for
$\eta\gg\max(1,\zeta)$. Let us analyze the remaining regions $\eta<1$ and $\zeta>\eta>1$.

The region $\zeta>\eta>1$ corresponds in domain 1 to 
\beq
\hat\omega\hat x_3&<
\frac{\hat k_\perp \hat x_3}{\kappa}\ln(\ex\hat x_3)<\frac{1}{\kappa}\ln(\ex\hat x_3)
\nonumber
\\
&\Rightarrow
1>\hat k\hat x_3>\frac{\kappa\hat\omega x_3}{\ln(\ex\hat x_3)}
\nonumber
\\
&\Rightarrow\hat\omega<\frac{\ln(\ex\hat x_3)}{\kappa\hat x_3}<\frac{1}{\kappa},
\label{items1}
\eeq
where $\hat x_3=\bar x_3/z_0$.
The region $\eta<1$ with $\hat x_3>1$, in turn, corresponds to $\hat\omega<1$. We thus reach
the conclusion that for $\hat\omega$ sufficiently large ($\hat \omega \gtrsim 2.5$),
integral range vortices do not contribute to $\hat S_g$.

\subsection{Large frequency limit}
The large $\hat\omega$ limit confines us to domain 2.
 The region $1<\eta<\zeta$ corresponds to
\beq
\hat\omega&<
\frac{p\hat k}{\kappa}\ln(\ex\hat x_3)<
\frac{\hat k}{\kappa}\ln(\ex\hat x_3)
\nonumber
\\
&\Rightarrow\hat k>\frac{\kappa\hat\omega}{\ln(\ex\hat x_3)},
\label{cond1bis}
\eeq
and the integral in $p$ in \cref{S_g PBL} is bounded in this range by
\beq
p>p_{\rm min}=
\frac{\kappa}{\hat k^{2/3}\ln(\ex\hat x_3)}.
\label{condpbis}
\eeq
Now, unless the detector is at the surface, the term $\hat r_0$ is  large, and thus,
unless $p\simeq 0$, the factor $\ex^{-2p\hat k\hat r_0}$ in \cref{S_g PBL}
is going to be
very small. The contribution to $\hat S_g$ from the region $\zeta>\eta>1$ is in this case
exponentially damped.

To get $p_{\rm min}=0$ we need to go to the region $\eta<1$, corresponding to the
condition on $k$:
\beq
\hat k>\hat k_{\rm min}\simeq \hat\omega^{3/2}\hat x_3^{1/2}.
\eeq
The function $B$ \eqref{B} is $O(1)$ in the whole
integration domain of
\cref{S_g PBL}, and we thus get
\begin{widetext}
\beq
\hat S_g\sim \int_1^{+\infty}\hat x_3^{-4/3}\d\hat x_3
\int_{\hat k_{\rm min}}^{+\infty}\hat k^{-13/3}\d\hat k\int_0^1p\,\d p\
\ex^{-2p\hat k\hat r_0}
\sim\hat\omega^{-8}\hat r_0^{-2}.
\label{large omega 1}
\eeq
\end{widetext}

\subsection{Small frequency limit}
Let us consider first domain 1.
The two conditions $\eta<1$ and $\bar x_3<\LO$ imply
\beq
\hat x_3<\hat x_{3,\rm{max}}=\min(\hat\omega^{-1},\hat L_{\rm O}).
\eeq
The contribution to $\hat S_g$ from domain 1 is therefore
\beq
\hat S_{g1}\sim\int_{\hat x_{3, \rm min}}^{\hat x_{3,\rm{max}}}\d\hat x_3\int_0^{1/\hat x_3}
\d\hat k\int_0^1\d p\ g_1(p,\hat k,\hat x_3),
\nonumber
\eeq
where $g_1(p,\hat k,\hat x_3)=pB\hat x^{3}\ex^{-2p\hat k\hat r_0}$. The integral is concentrated at
$(\hat x_3,\hat k)\sim (\hat x_{3,\rm{max}},0)$. We find the limit behaviors
\beq
\hat S_{g1}\sim \begin{cases}
\hat x^3_{3, \rm max},& \hat r_0\lesssim \hat x_{3,\rm{max}}
\\
\hat x_{3, \rm max}^5\hat r_0^{-2}, & \hat r_0 \gg x_{3,\rm{max}}
\end{cases}
\label{smallish omega}
\eeq
Let us switch to domain 2 and continue to focus on the region $\eta<1$, for
which $A\sim 1$. We have now
\beq
\hat S_{g2}\sim\int_{\hat x_{3, \rm min}}^{\hat x_{3,\rm{max}}}\d\hat x_3\int_{1/\hat x_3}^{+\infty}
\d\hat k\int_0^1\d p\ g_2(p,\hat k,\hat x_3),
\nonumber
\eeq
where $g_2(p,\hat k,\hat x_3)=pB\hat k^{-13/3}\hat x_3^{-4/3}\ex^{-2p\hat k\hat r_0}$. In this case the integral is concentrated at $(\hat x_3,\hat k)\sim
(\hat x_{3,\rm{max}},1/\hat x_{3,\rm{max}})$. We can verify that $\hat S_{g2}\sim \hat S_{g1}$, and thus recover \cref{small omega,x3max}.

\section{General scaling of the noise spectra with $r_0$}
\label{sec:Scalingr0}

In \cref{sectNNhom}  and \cref{sec:WallTurbulence}  we have shown that, in some limiting cases $S_g$ scales as $1/r_0^2$  (see \cref{weak wind,large omega,small omega}). Here we prove that such a behavior generally arises whenever vortex decay dominates over wind advection  and the detector depth is sufficiently large, independently of the chosen turbulence model. From \cref{int,S_g PBL}, we  can  see that $S_g$ can be expressed in the general form
\beq\label{eq:general_form}
    S_g&\sim\int_0^1\d p\frac{p}{\sqrt{1-p^2}}\int_{x_\text{min}}^L\d x\int_0^{2\pi}\d\phi\cos^2\phi\\
    &\times\int_0^{\infty} \d k f(p,x,\phi,k)h[\tau_k(\omega-p\cos\phi k U)] F(pkr_0).\nonumber
\eeq
Here $f$ is the spatial part of the correlation functions, $h$ contains information about time correlations and $F$ is a function whose form depend on the chosen geometry. In particular, $F(pkr_0)=\ex^{-2p k r_0}$ in our case. We assume that:
\begin{itemize}
    \item The integral \eqref{eq:general_form} is convergent, which is always the case for reasonable models;
    \item The function $h$ satisfies the properties given in \cref{sectNNhom} below \cref{TTh}, i.e. $h(z)$ has a maximum for $z=0$ and $h(z)\to0$ at least exponentially for $z\to\infty$;
    \item $f$ is regular over the whole integration domain;
    \item $F(pkr_0)\to 0$ for $pkr_0\gg1$ sufficiently fast (at least exponentially).
\end{itemize}
Due to the properties of $h$, we see that the contribution to the integral over $k$ in \cref{eq:general_form} will be peaked around some value $k=\bar{k}$, whose specific value depends on the other parameters and on the specific model. Thus
\beq
    S_g&\sim\int_0^1\d p\frac{p}{\sqrt{1-p^2}}\int_{x_\text{min}}^L\d x\int_0^{2\pi}\d\phi\cos^2\phi\\
    &\times\bar{k} f(p,x,\phi,\bar{k})h[\tau_{\bar{k}}(\omega-p\cos\phi \bar{k} U)] F(p\bar{k}r_0).\nonumber
\eeq
If we now assume $\omega\gg\bar{k} U$, i.e. vortex decay dominates over wind advection,  we see that the function $h$ will become independent of the wind speed. One can now performe the integrals over $x$ and $\phi$, so that
\beq
    S_g&\sim \bar{k}h(\tau_{\bar{k}}\omega)\int_0^1\d p\frac{p}{\sqrt{1-p^2}}\\
    &\times  g(p,x_\text{min},L,\bar{k})F(p\bar{k}r_0).\nonumber
\eeq
When $r_0\gg 1/\bar{k}$, we see that the only non-negligible contribution to the integral comes from the values of $p\lesssim 1/(\bar{k}r_0)\ll 1$, according to the last assumption above. The integral becomes then
\beq
    S_g&\sim \bar{k}h(\tau_{\bar{k}}\omega)g(0,x_\text{min},L,\bar{k})F(0)\int_0^{1/(\bar{k} r_0)}\d p\,p\\
    &\sim\frac{h(\tau_{\bar{k}}\omega)g(0,x_\text{min},L,\bar{k})F(0)}{\bar{k} r_0^2}.\nonumber
\eeq

\bibliography{refs}

\begin{thebibliography}{34}%
\makeatletter
\providecommand \@ifxundefined [1]{%
 \@ifx{#1\undefined}
}%
\providecommand \@ifnum [1]{%
 \ifnum #1\expandafter \@firstoftwo
 \else \expandafter \@secondoftwo
 \fi
}%
\providecommand \@ifx [1]{%
 \ifx #1\expandafter \@firstoftwo
 \else \expandafter \@secondoftwo
 \fi
}%
\providecommand \natexlab [1]{#1}%
\providecommand \enquote  [1]{``#1''}%
\providecommand \bibnamefont  [1]{#1}%
\providecommand \bibfnamefont [1]{#1}%
\providecommand \citenamefont [1]{#1}%
\providecommand \href@noop [0]{\@secondoftwo}%
\providecommand \href [0]{\begingroup \@sanitize@url \@href}%
\providecommand \@href[1]{\@@startlink{#1}\@@href}%
\providecommand \@@href[1]{\endgroup#1\@@endlink}%
\providecommand \@sanitize@url [0]{\catcode `\\12\catcode `\$12\catcode
  `\&12\catcode `\#12\catcode `\^12\catcode `\_12\catcode `\%12\relax}%
\providecommand \@@startlink[1]{}%
\providecommand \@@endlink[0]{}%
\providecommand \url  [0]{\begingroup\@sanitize@url \@url }%
\providecommand \@url [1]{\endgroup\@href {#1}{\urlprefix }}%
\providecommand \urlprefix  [0]{URL }%
\providecommand \Eprint [0]{\href }%
\providecommand \doibase [0]{http://dx.doi.org/}%
\providecommand \selectlanguage [0]{\@gobble}%
\providecommand \bibinfo  [0]{\@secondoftwo}%
\providecommand \bibfield  [0]{\@secondoftwo}%
\providecommand \translation [1]{[#1]}%
\providecommand \BibitemOpen [0]{}%
\providecommand \bibitemStop [0]{}%
\providecommand \bibitemNoStop [0]{.\EOS\space}%
\providecommand \EOS [0]{\spacefactor3000\relax}%
\providecommand \BibitemShut  [1]{\csname bibitem#1\endcsname}%
\let\auto@bib@innerbib\@empty
%</preamble>
\bibitem [{\citenamefont {Abbott}\ \emph
  {et~al.}(2016{\natexlab{a}})\citenamefont {Abbott} \emph
  {et~al.}}]{LIGOScientific:2016aoc}%
  \BibitemOpen
  \bibfield  {author} {\bibinfo {author} {\bibfnamefont {B.~P.}\ \bibnamefont
  {Abbott}} \emph {et~al.} (\bibinfo {collaboration} {LIGO Scientific,
  Virgo}),\ }\href {\doibase 10.1103/PhysRevLett.116.061102} {\bibfield
  {journal} {\bibinfo  {journal} {Phys. Rev. Lett.}\ }\textbf {\bibinfo
  {volume} {116}},\ \bibinfo {pages} {061102} (\bibinfo {year}
  {2016}{\natexlab{a}})},\ \Eprint {http://arxiv.org/abs/1602.03837}
  {arXiv:1602.03837 [gr-qc]} \BibitemShut {NoStop}%
\bibitem [{\citenamefont {Acernese}\ \emph {et~al.}(2015)\citenamefont
  {Acernese} \emph {et~al.}}]{VIRGO:2014yos}%
  \BibitemOpen
  \bibfield  {author} {\bibinfo {author} {\bibfnamefont {F.}~\bibnamefont
  {Acernese}} \emph {et~al.} (\bibinfo {collaboration} {VIRGO}),\ }\href
  {\doibase 10.1088/0264-9381/32/2/024001} {\bibfield  {journal} {\bibinfo
  {journal} {Class. Quant. Grav.}\ }\textbf {\bibinfo {volume} {32}},\ \bibinfo
  {pages} {024001} (\bibinfo {year} {2015})},\ \Eprint
  {http://arxiv.org/abs/1408.3978} {arXiv:1408.3978 [gr-qc]} \BibitemShut
  {NoStop}%
\bibitem [{\citenamefont {Abbott}\ \emph
  {et~al.}(2016{\natexlab{b}})\citenamefont {Abbott} \emph
  {et~al.}}]{LIGOScientific:2016sjg}%
  \BibitemOpen
  \bibfield  {author} {\bibinfo {author} {\bibfnamefont {B.~P.}\ \bibnamefont
  {Abbott}} \emph {et~al.} (\bibinfo {collaboration} {LIGO Scientific,
  Virgo}),\ }\href {\doibase 10.1103/PhysRevLett.116.241103} {\bibfield
  {journal} {\bibinfo  {journal} {Phys. Rev. Lett.}\ }\textbf {\bibinfo
  {volume} {116}},\ \bibinfo {pages} {241103} (\bibinfo {year}
  {2016}{\natexlab{b}})},\ \Eprint {http://arxiv.org/abs/1606.04855}
  {arXiv:1606.04855 [gr-qc]} \BibitemShut {NoStop}%
\bibitem [{\citenamefont {Abbott}\ \emph
  {et~al.}(2017{\natexlab{a}})\citenamefont {Abbott} \emph
  {et~al.}}]{LIGOScientific:2017vwq}%
  \BibitemOpen
  \bibfield  {author} {\bibinfo {author} {\bibfnamefont {B.~P.}\ \bibnamefont
  {Abbott}} \emph {et~al.} (\bibinfo {collaboration} {LIGO Scientific,
  Virgo}),\ }\href {\doibase 10.1103/PhysRevLett.119.161101} {\bibfield
  {journal} {\bibinfo  {journal} {Phys. Rev. Lett.}\ }\textbf {\bibinfo
  {volume} {119}},\ \bibinfo {pages} {161101} (\bibinfo {year}
  {2017}{\natexlab{a}})},\ \Eprint {http://arxiv.org/abs/1710.05832}
  {arXiv:1710.05832 [gr-qc]} \BibitemShut {NoStop}%
\bibitem [{\citenamefont {Abbott}\ \emph
  {et~al.}(2017{\natexlab{b}})\citenamefont {Abbott} \emph
  {et~al.}}]{LIGOScientific:2017ync}%
  \BibitemOpen
  \bibfield  {author} {\bibinfo {author} {\bibfnamefont {B.~P.}\ \bibnamefont
  {Abbott}} \emph {et~al.} (\bibinfo {collaboration} {LIGO Scientific, Virgo,
  Fermi GBM, INTEGRAL, IceCube, AstroSat Cadmium Zinc Telluride Imager Team,
  IPN, Insight-Hxmt, ANTARES, Swift, AGILE Team, 1M2H Team, Dark Energy Camera
  GW-EM, DES, DLT40, GRAWITA, Fermi-LAT, ATCA, ASKAP, Las Cumbres Observatory
  Group, OzGrav, DWF (Deeper Wider Faster Program), AST3, CAASTRO, VINROUGE,
  MASTER, J-GEM, GROWTH, JAGWAR, CaltechNRAO, TTU-NRAO, NuSTAR, Pan-STARRS,
  MAXI Team, TZAC Consortium, KU, Nordic Optical Telescope, ePESSTO, GROND,
  Texas Tech University, SALT Group, TOROS, BOOTES, MWA, CALET, IKI-GW
  Follow-up, H.E.S.S., LOFAR, LWA, HAWC, Pierre Auger, ALMA, Euro VLBI Team, Pi
  of Sky, Chandra Team at McGill University, DFN, ATLAS Telescopes, High Time
  Resolution Universe Survey, RIMAS, RATIR, SKA South Africa/MeerKAT}),\ }\href
  {\doibase 10.3847/2041-8213/aa91c9} {\bibfield  {journal} {\bibinfo
  {journal} {Astrophys. J. Lett.}\ }\textbf {\bibinfo {volume} {848}},\
  \bibinfo {pages} {L12} (\bibinfo {year} {2017}{\natexlab{b}})},\ \Eprint
  {http://arxiv.org/abs/1710.05833} {arXiv:1710.05833 [astro-ph.HE]}
  \BibitemShut {NoStop}%
\bibitem [{\citenamefont {Abbott}\ \emph
  {et~al.}(2017{\natexlab{c}})\citenamefont {Abbott} \emph
  {et~al.}}]{LIGOScientific:2017zic}%
  \BibitemOpen
  \bibfield  {author} {\bibinfo {author} {\bibfnamefont {B.~P.}\ \bibnamefont
  {Abbott}} \emph {et~al.} (\bibinfo {collaboration} {LIGO Scientific, Virgo,
  Fermi-GBM, INTEGRAL}),\ }\href {\doibase 10.3847/2041-8213/aa920c} {\bibfield
   {journal} {\bibinfo  {journal} {Astrophys. J. Lett.}\ }\textbf {\bibinfo
  {volume} {848}},\ \bibinfo {pages} {L13} (\bibinfo {year}
  {2017}{\natexlab{c}})},\ \Eprint {http://arxiv.org/abs/1710.05834}
  {arXiv:1710.05834 [astro-ph.HE]} \BibitemShut {NoStop}%
\bibitem [{\citenamefont {Abbott}\ \emph
  {et~al.}(2016{\natexlab{c}})\citenamefont {Abbott} \emph
  {et~al.}}]{LIGOScientific:2016lio}%
  \BibitemOpen
  \bibfield  {author} {\bibinfo {author} {\bibfnamefont {B.~P.}\ \bibnamefont
  {Abbott}} \emph {et~al.} (\bibinfo {collaboration} {LIGO Scientific,
  Virgo}),\ }\href {\doibase 10.1103/PhysRevLett.116.221101} {\bibfield
  {journal} {\bibinfo  {journal} {Phys. Rev. Lett.}\ }\textbf {\bibinfo
  {volume} {116}},\ \bibinfo {pages} {221101} (\bibinfo {year}
  {2016}{\natexlab{c}})},\ \bibinfo {note} {[Erratum: Phys.Rev.Lett. 121,
  129902 (2018)]},\ \Eprint {http://arxiv.org/abs/1602.03841} {arXiv:1602.03841
  [gr-qc]} \BibitemShut {NoStop}%
\bibitem [{\citenamefont {Abbott}\ \emph {et~al.}(2019)\citenamefont {Abbott}
  \emph {et~al.}}]{LIGOScientific:2019fpa}%
  \BibitemOpen
  \bibfield  {author} {\bibinfo {author} {\bibfnamefont {B.~P.}\ \bibnamefont
  {Abbott}} \emph {et~al.} (\bibinfo {collaboration} {LIGO Scientific,
  Virgo}),\ }\href {\doibase 10.1103/PhysRevD.100.104036} {\bibfield  {journal}
  {\bibinfo  {journal} {Phys. Rev. D}\ }\textbf {\bibinfo {volume} {100}},\
  \bibinfo {pages} {104036} (\bibinfo {year} {2019})},\ \Eprint
  {http://arxiv.org/abs/1903.04467} {arXiv:1903.04467 [gr-qc]} \BibitemShut
  {NoStop}%
\bibitem [{\citenamefont {Abbott}\ \emph
  {et~al.}(2021{\natexlab{a}})\citenamefont {Abbott} \emph
  {et~al.}}]{LIGOScientific:2020ibl}%
  \BibitemOpen
  \bibfield  {author} {\bibinfo {author} {\bibfnamefont {R.}~\bibnamefont
  {Abbott}} \emph {et~al.} (\bibinfo {collaboration} {LIGO Scientific,
  Virgo}),\ }\href {\doibase 10.1103/PhysRevX.11.021053} {\bibfield  {journal}
  {\bibinfo  {journal} {Phys. Rev. X}\ }\textbf {\bibinfo {volume} {11}},\
  \bibinfo {pages} {021053} (\bibinfo {year} {2021}{\natexlab{a}})},\ \Eprint
  {http://arxiv.org/abs/2010.14527} {arXiv:2010.14527 [gr-qc]} \BibitemShut
  {NoStop}%
\bibitem [{\citenamefont {Abbott}\ \emph
  {et~al.}(2021{\natexlab{b}})\citenamefont {Abbott} \emph
  {et~al.}}]{LIGOScientific:2021qlt}%
  \BibitemOpen
  \bibfield  {author} {\bibinfo {author} {\bibfnamefont {R.}~\bibnamefont
  {Abbott}} \emph {et~al.} (\bibinfo {collaboration} {LIGO Scientific, KAGRA,
  VIRGO}),\ }\href {\doibase 10.3847/2041-8213/ac082e} {\bibfield  {journal}
  {\bibinfo  {journal} {Astrophys. J. Lett.}\ }\textbf {\bibinfo {volume}
  {915}},\ \bibinfo {pages} {L5} (\bibinfo {year} {2021}{\natexlab{b}})},\
  \Eprint {http://arxiv.org/abs/2106.15163} {arXiv:2106.15163 [astro-ph.HE]}
  \BibitemShut {NoStop}%
\bibitem [{\citenamefont {Somiya}(2012)}]{Somiya:2011np}%
  \BibitemOpen
  \bibfield  {author} {\bibinfo {author} {\bibfnamefont {K.}~\bibnamefont
  {Somiya}} (\bibinfo {collaboration} {KAGRA}),\ }\href {\doibase
  10.1088/0264-9381/29/12/124007} {\bibfield  {journal} {\bibinfo  {journal}
  {Class. Quant. Grav.}\ }\textbf {\bibinfo {volume} {29}},\ \bibinfo {pages}
  {124007} (\bibinfo {year} {2012})},\ \Eprint {http://arxiv.org/abs/1111.7185}
  {arXiv:1111.7185 [gr-qc]} \BibitemShut {NoStop}%
\bibitem [{\citenamefont {Abbott}\ \emph {et~al.}(2018)\citenamefont {Abbott}
  \emph {et~al.}}]{KAGRA:2013rdx}%
  \BibitemOpen
  \bibfield  {author} {\bibinfo {author} {\bibfnamefont {B.~P.}\ \bibnamefont
  {Abbott}} \emph {et~al.} (\bibinfo {collaboration} {KAGRA, LIGO Scientific,
  Virgo, VIRGO}),\ }\href {\doibase 10.1007/s41114-020-00026-9} {\bibfield
  {journal} {\bibinfo  {journal} {Living Rev. Rel.}\ }\textbf {\bibinfo
  {volume} {21}},\ \bibinfo {pages} {3} (\bibinfo {year} {2018})},\ \Eprint
  {http://arxiv.org/abs/1304.0670} {arXiv:1304.0670 [gr-qc]} \BibitemShut
  {NoStop}%
\bibitem [{\citenamefont {Akutsu}\ \emph {et~al.}(2021)\citenamefont {Akutsu}
  \emph {et~al.}}]{KAGRA:2020tym}%
  \BibitemOpen
  \bibfield  {author} {\bibinfo {author} {\bibfnamefont {T.}~\bibnamefont
  {Akutsu}} \emph {et~al.} (\bibinfo {collaboration} {KAGRA}),\ }\href
  {\doibase 10.1093/ptep/ptaa125} {\bibfield  {journal} {\bibinfo  {journal}
  {PTEP}\ }\textbf {\bibinfo {volume} {2021}},\ \bibinfo {pages} {05A101}
  (\bibinfo {year} {2021})},\ \Eprint {http://arxiv.org/abs/2005.05574}
  {arXiv:2005.05574 [physics.ins-det]} \BibitemShut {NoStop}%
\bibitem [{\citenamefont {Punturo}\ \emph {et~al.}(2010)\citenamefont {Punturo}
  \emph {et~al.}}]{Punturo:2010zz}%
  \BibitemOpen
  \bibfield  {author} {\bibinfo {author} {\bibfnamefont {M.}~\bibnamefont
  {Punturo}} \emph {et~al.},\ }\href {\doibase 10.1088/0264-9381/27/19/194002}
  {\bibfield  {journal} {\bibinfo  {journal} {Class. Quant. Grav.}\ }\textbf
  {\bibinfo {volume} {27}},\ \bibinfo {pages} {194002} (\bibinfo {year}
  {2010})}\BibitemShut {NoStop}%
\bibitem [{\citenamefont {Abbott}\ \emph
  {et~al.}(2017{\natexlab{d}})\citenamefont {Abbott} \emph
  {et~al.}}]{LIGOScientific:2016wof}%
  \BibitemOpen
  \bibfield  {author} {\bibinfo {author} {\bibfnamefont {B.~P.}\ \bibnamefont
  {Abbott}} \emph {et~al.} (\bibinfo {collaboration} {LIGO Scientific}),\
  }\href {\doibase 10.1088/1361-6382/aa51f4} {\bibfield  {journal} {\bibinfo
  {journal} {Class. Quant. Grav.}\ }\textbf {\bibinfo {volume} {34}},\ \bibinfo
  {pages} {044001} (\bibinfo {year} {2017}{\natexlab{d}})},\ \Eprint
  {http://arxiv.org/abs/1607.08697} {arXiv:1607.08697 [astro-ph.IM]}
  \BibitemShut {NoStop}%
\bibitem [{\citenamefont {Sathyaprakash}\ \emph {et~al.}(2012)\citenamefont
  {Sathyaprakash} \emph {et~al.}}]{Sathyaprakash:2012jk}%
  \BibitemOpen
  \bibfield  {author} {\bibinfo {author} {\bibfnamefont {B.}~\bibnamefont
  {Sathyaprakash}} \emph {et~al.},\ }\href {\doibase
  10.1088/0264-9381/29/12/124013} {\bibfield  {journal} {\bibinfo  {journal}
  {Class. Quant. Grav.}\ }\textbf {\bibinfo {volume} {29}},\ \bibinfo {pages}
  {124013} (\bibinfo {year} {2012})},\ \bibinfo {note} {[Erratum:
  Class.Quant.Grav. 30, 079501 (2013)]},\ \Eprint
  {http://arxiv.org/abs/1206.0331} {arXiv:1206.0331 [gr-qc]} \BibitemShut
  {NoStop}%
\bibitem [{\citenamefont {Maggiore}\ \emph {et~al.}(2020)\citenamefont
  {Maggiore} \emph {et~al.}}]{Maggiore:2019uih}%
  \BibitemOpen
  \bibfield  {author} {\bibinfo {author} {\bibfnamefont {M.}~\bibnamefont
  {Maggiore}} \emph {et~al.},\ }\href {\doibase 10.1088/1475-7516/2020/03/050}
  {\bibfield  {journal} {\bibinfo  {journal} {JCAP}\ }\textbf {\bibinfo
  {volume} {03}},\ \bibinfo {pages} {050} (\bibinfo {year} {2020})},\ \Eprint
  {http://arxiv.org/abs/1912.02622} {arXiv:1912.02622 [astro-ph.CO]}
  \BibitemShut {NoStop}%
\bibitem [{\citenamefont {Harms}(2019)}]{Har2019}%
  \BibitemOpen
  \bibfield  {author} {\bibinfo {author} {\bibfnamefont {J.}~\bibnamefont
  {Harms}},\ }\href {\doibase 10.1007/s41114-019-0022-2} {\bibfield  {journal}
  {\bibinfo  {journal} {Living Reviews in Relativity}\ }\textbf {\bibinfo
  {volume} {22}},\ \bibinfo {pages} {6} (\bibinfo {year} {2019})}\BibitemShut
  {NoStop}%
\bibitem [{\citenamefont {Harms}\ \emph {et~al.}(2022)\citenamefont {Harms},
  \citenamefont {Naticchioni}, \citenamefont {Calloni}, \citenamefont
  {De~Rosa}, \citenamefont {Ricci},\ and\ \citenamefont
  {D'Urso}}]{Harms:2022jth}%
  \BibitemOpen
  \bibfield  {author} {\bibinfo {author} {\bibfnamefont {J.}~\bibnamefont
  {Harms}}, \bibinfo {author} {\bibfnamefont {L.}~\bibnamefont {Naticchioni}},
  \bibinfo {author} {\bibfnamefont {E.}~\bibnamefont {Calloni}}, \bibinfo
  {author} {\bibfnamefont {R.}~\bibnamefont {De~Rosa}}, \bibinfo {author}
  {\bibfnamefont {F.}~\bibnamefont {Ricci}}, \ and\ \bibinfo {author}
  {\bibfnamefont {D.}~\bibnamefont {D'Urso}},\ }\href@noop {} {\  (\bibinfo
  {year} {2022})},\ \Eprint {http://arxiv.org/abs/2202.12841} {arXiv:2202.12841
  [gr-qc]} \BibitemShut {NoStop}%
\bibitem [{\citenamefont {Saulson}(1984)}]{Saulson:1984yg}%
  \BibitemOpen
  \bibfield  {author} {\bibinfo {author} {\bibfnamefont {P.~R.}\ \bibnamefont
  {Saulson}},\ }\href {\doibase 10.1103/PhysRevD.30.732} {\bibfield  {journal}
  {\bibinfo  {journal} {Phys. Rev. D}\ }\textbf {\bibinfo {volume} {30}},\
  \bibinfo {pages} {732} (\bibinfo {year} {1984})}\BibitemShut {NoStop}%
\bibitem [{\citenamefont {Creighton}(2008)}]{Creighton:2000gu}%
  \BibitemOpen
  \bibfield  {author} {\bibinfo {author} {\bibfnamefont {T.}~\bibnamefont
  {Creighton}},\ }\href {\doibase 10.1088/0264-9381/25/12/125011} {\bibfield
  {journal} {\bibinfo  {journal} {Class. Quant. Grav.}\ }\textbf {\bibinfo
  {volume} {25}},\ \bibinfo {pages} {125011} (\bibinfo {year} {2008})},\
  \Eprint {http://arxiv.org/abs/gr-qc/0007050} {arXiv:gr-qc/0007050}
  \BibitemShut {NoStop}%
\bibitem [{\citenamefont {Cafaro}\ and\ \citenamefont
  {Ali}(2009)}]{Cafaro:2009mu}%
  \BibitemOpen
  \bibfield  {author} {\bibinfo {author} {\bibfnamefont {C.}~\bibnamefont
  {Cafaro}}\ and\ \bibinfo {author} {\bibfnamefont {S.~A.}\ \bibnamefont
  {Ali}},\ }\href@noop {} {\  (\bibinfo {year} {2009})},\ \Eprint
  {http://arxiv.org/abs/0906.4844} {arXiv:0906.4844 [gr-qc]} \BibitemShut
  {NoStop}%
\bibitem [{\citenamefont {Stull}(1988)}]{stull1988introduction}%
  \BibitemOpen
  \bibfield  {author} {\bibinfo {author} {\bibfnamefont {R.~B.}\ \bibnamefont
  {Stull}},\ }\href@noop {} {\emph {\bibinfo {title} {An introduction to
  boundary layer meteorology}}},\ Vol.~\bibinfo {volume} {13}\ (\bibinfo
  {publisher} {Springer Science \& Business Media},\ \bibinfo {year}
  {1988})\BibitemShut {NoStop}%
\bibitem [{\citenamefont {Schlichting}\ and\ \citenamefont
  {Gersten}(2003)}]{schlichting2003boundary}%
  \BibitemOpen
  \bibfield  {author} {\bibinfo {author} {\bibfnamefont {H.}~\bibnamefont
  {Schlichting}}\ and\ \bibinfo {author} {\bibfnamefont {K.}~\bibnamefont
  {Gersten}},\ }\href@noop {} {\emph {\bibinfo {title} {Boundary-layer
  theory}}}\ (\bibinfo  {publisher} {Springer Science \& Business Media},\
  \bibinfo {year} {2003})\BibitemShut {NoStop}%
\bibitem [{\citenamefont {Troen}\ and\ \citenamefont {{Lundtang
  Petersen}}(1989)}]{troen1989european}%
  \BibitemOpen
  \bibfield  {author} {\bibinfo {author} {\bibfnamefont {I.}~\bibnamefont
  {Troen}}\ and\ \bibinfo {author} {\bibfnamefont {E.}~\bibnamefont {{Lundtang
  Petersen}}},\ }\href@noop {} {{\selectlanguage {English}\emph {\bibinfo
  {title} {European Wind Atlas}}}}\ (\bibinfo  {publisher} {Ris{\o} National
  Laboratory},\ \bibinfo {year} {1989})\BibitemShut {NoStop}%
\bibitem [{\citenamefont {Obukhov}(1971)}]{obukhov1971turbulence}%
  \BibitemOpen
  \bibfield  {author} {\bibinfo {author} {\bibfnamefont {A.}~\bibnamefont
  {Obukhov}},\ }\href {\doibase 10.1007/BF00718085} {\bibfield  {journal}
  {\bibinfo  {journal} {Boundary-Layer Meteorology}\ }\textbf {\bibinfo
  {volume} {2}},\ \bibinfo {pages} {7} (\bibinfo {year} {1971})}\BibitemShut
  {NoStop}%
\bibitem [{\citenamefont {Kolmogorov}(1991)}]{kolmogorov1991local}%
  \BibitemOpen
  \bibfield  {author} {\bibinfo {author} {\bibfnamefont {A.~N.}\ \bibnamefont
  {Kolmogorov}},\ }\href {\doibase 10.1098/rspa.1991.0075} {\bibfield
  {journal} {\bibinfo  {journal} {Proceedings of the Royal Society of London.
  Series A: Mathematical and Physical Sciences}\ }\textbf {\bibinfo {volume}
  {434}},\ \bibinfo {pages} {9} (\bibinfo {year} {1991})}\BibitemShut {NoStop}%
\bibitem [{\citenamefont {Belinicher}\ and\ \citenamefont
  {L’vov}(1987)}]{belinicher1987scale}%
  \BibitemOpen
  \bibfield  {author} {\bibinfo {author} {\bibfnamefont {V.}~\bibnamefont
  {Belinicher}}\ and\ \bibinfo {author} {\bibfnamefont {V.}~\bibnamefont
  {L’vov}},\ }\href {\doibase 10.1016/0370-1573(91)90081-V} {\bibfield
  {journal} {\bibinfo  {journal} {Sov. Phys. JETP}\ }\textbf {\bibinfo {volume}
  {66}},\ \bibinfo {pages} {303} (\bibinfo {year} {1987})}\BibitemShut
  {NoStop}%
\bibitem [{\citenamefont {Wu}\ and\ \citenamefont
  {He}(2021)}]{wu2021stochastic}%
  \BibitemOpen
  \bibfield  {author} {\bibinfo {author} {\bibfnamefont {T.}~\bibnamefont
  {Wu}}\ and\ \bibinfo {author} {\bibfnamefont {G.}~\bibnamefont {He}},\ }\href
  {\doibase 10.1103/PhysRevFluids.6.054602} {\bibfield  {journal} {\bibinfo
  {journal} {Physical Review Fluids}\ }\textbf {\bibinfo {volume} {6}},\
  \bibinfo {pages} {054602} (\bibinfo {year} {2021})}\BibitemShut {NoStop}%
\bibitem [{\citenamefont {Naguib}\ and\ \citenamefont
  {Wark}(1992)}]{naguib1992investigation}%
  \BibitemOpen
  \bibfield  {author} {\bibinfo {author} {\bibfnamefont {A.}~\bibnamefont
  {Naguib}}\ and\ \bibinfo {author} {\bibfnamefont {C.}~\bibnamefont {Wark}},\
  }\href {\doibase 10.1017/S0022112092002817} {\bibfield  {journal} {\bibinfo
  {journal} {Journal of Fluid Mechanics}\ }\textbf {\bibinfo {volume} {243}},\
  \bibinfo {pages} {541} (\bibinfo {year} {1992})}\BibitemShut {NoStop}%
\bibitem [{GIT(2022)}]{GITHub_repo}%
  \BibitemOpen
  \href@noop {} {\enquote {\bibinfo {title} {{\texttt{AtmosphericNN} GitHub
  repository}},}\ }\bibinfo {howpublished}
  {\url{https://github.com/maurooi/AtmosphericNN}} (\bibinfo {year}
  {2022})\BibitemShut {NoStop}%
\bibitem [{\citenamefont {{Peter Lepage}}(1978)}]{Vegas}%
  \BibitemOpen
  \bibfield  {author} {\bibinfo {author} {\bibfnamefont {G.}~\bibnamefont
  {{Peter Lepage}}},\ }\href {\doibase
  https://doi.org/10.1016/0021-9991(78)90004-9} {\bibfield  {journal} {\bibinfo
   {journal} {Journal of Computational Physics}\ }\textbf {\bibinfo {volume}
  {27}},\ \bibinfo {pages} {192} (\bibinfo {year} {1978})}\BibitemShut
  {NoStop}%
\bibitem [{\citenamefont {Alves~Junior}(2018)}]{hydra}%
  \BibitemOpen
  \bibfield  {author} {\bibinfo {author} {\bibfnamefont {A.~A.}\ \bibnamefont
  {Alves~Junior}},\ }\href {\doibase 10.5281/zenodo.1206261} {\enquote
  {\bibinfo {title} {Multithreadcorner/hydra},}\ } (\bibinfo {year}
  {2018})\BibitemShut {NoStop}%
\bibitem [{\citenamefont {Stoll}\ \emph {et~al.}(2020)\citenamefont {Stoll},
  \citenamefont {Gibbs}, \citenamefont {Salesky}, \citenamefont {Anderson},\
  and\ \citenamefont {Calaf}}]{stoll2020large}%
  \BibitemOpen
  \bibfield  {author} {\bibinfo {author} {\bibfnamefont {R.}~\bibnamefont
  {Stoll}}, \bibinfo {author} {\bibfnamefont {J.~A.}\ \bibnamefont {Gibbs}},
  \bibinfo {author} {\bibfnamefont {S.~T.}\ \bibnamefont {Salesky}}, \bibinfo
  {author} {\bibfnamefont {W.}~\bibnamefont {Anderson}}, \ and\ \bibinfo
  {author} {\bibfnamefont {M.}~\bibnamefont {Calaf}},\ }\href {\doibase
  10.1007/s10546-020-00556-3} {\bibfield  {journal} {\bibinfo  {journal}
  {Boundary-Layer Meteorology}\ }\textbf {\bibinfo {volume} {177}},\ \bibinfo
  {pages} {541} (\bibinfo {year} {2020})}\BibitemShut {NoStop}%
\end{thebibliography}%
\end{document}